\documentclass[twocolumn,showpacs,preprintnumbers,amsmath,amssymb,prl]{revtex4}

\usepackage{graphicx}
\usepackage{bm}

\begin{document}


\title{Core-Softened System With Attraction: Trajectory Dependence of Anomalous Behavior}

\author{Yu. D. Fomin}
\affiliation{Institute for High Pressure Physics, Russian Academy
of Sciences, Troitsk 142190, Moscow Region, Russia}

\author{V. N. Ryzhov}
\affiliation{Institute for High Pressure Physics, Russian Academy
of Sciences, Troitsk 142190, Moscow Region, Russia and Moscow
Institute of Physics and Technology, Dolgoprudnii, Moscow region,
Russia}

\author{E. N. Tsiok}
\affiliation{Institute for High Pressure Physics, Russian Academy
of Sciences, Troitsk 142190, Moscow Region, Russia}

\date{\today}

\begin{abstract}
In the present article we carry out a molecular dynamics study of
the core-softened system and show that the existence of the
water-like anomalies in this system depends on the trajectory in
$P-\rho-T$ space along which the behavior of the system is
studied. For example, diffusion and structural anomalies are
visible along isotherms as a function of density, but disappears
along the isochores and isobars as a function of temperature. On
the other hand, the diffusion anomaly may be seen along adiabats
as a function of temperature, density and pressure. It should be
noted that it may be no signature of a particular anomaly along a
particular trajectory, but the anomalous region for that
particular anomaly can be defined when all possible trajectories
in the same space are examined (for example, signature of
diffusion anomaly is evident through the crossing of different
isochors. However, there is no signature of diffusion anomaly
along a particular isochor). We also analyze the applicability of
the Rosenfeld entropy scaling relations to this system in the
regions with the water-like anomalies. It is shown that the
validity of the Rosenfeld scaling relation for the diffusion
coefficient also depends on the trajectory in the $P-\rho-T$ space
along which the kinetic coefficients and the excess entropy are
calculated.
\end{abstract}

 \maketitle

\section{I. Introduction}

It is well known that some liquids (for example, water, silica,
silicon, carbon, phosphorus, and some biological systems) show an
anomalous behavior in the vicinity of their freezing lines
\cite{deben2003,bul2002,
angel2004,book,book1,deben2001,netz,stanley1,stanley2,ad1,ad2,ad3,ad4,ad5,
ad6,ad7,ad8,errington,errington2,st_bio,bagchi1}. The water phase
diagrams have regions where a thermal expansion coefficient is
negative (density anomaly), self-diffusivity increases upon
compression (diffusion anomaly), and the structural order of the
system decreases with increasing pressure (structural anomaly)
\cite{deben2001,netz}.

The first anomaly mentioned above is density anomaly. It means
that density increases upon heating or that the thermal expansion
coefficient becomes negative. Using the thermodynamic relation
$\left(\partial P/\partial T\right)_V=\alpha_P/K_T$, where
$\alpha_P$ is a thermal expansion coefficient and $K_T$ is the
isothermal compressibility and taking into account that $K_T$ is
always positive and finite for systems in equilibrium not at a
critical point, we conclude that density anomaly corresponds to
minimum of the pressure dependence on temperature along an
isochor. This is the most convenient indicator of density anomaly
in computer simulation.

If we consider a simple liquid (for, example, Lennard-Jones
liquid), and trace the diffusion along an isotherm we find that
the diffusion decreases under densification. This observation is
intuitively clear - if density increases the free volume decreases
and the particles have less freedom to move. However, some
substances have a region in density - temperature plane where
diffusion grows under densification. This is called anomalous
diffusion region which reflects the contradiction of this behavior
with the free volume picture described above. This means that
diffusion anomaly involves more complex mechanisms which will be
discussed below.

The last anomaly we discuss here is structural anomaly. Initially
this anomaly was introduced via order parameters characterizing
the local order in liquid. However, later on the local order was
also related to excess entropy of the liquid which is defined as
the difference between the entropy and the ideal gas entropy at
the same $(\rho,T)$ point: $S_{ex}=S-S_{id}$. In normal liquid
excess entropy is monotonically decaying function of density along
an isotherm while in anomalous liquids it demonstrates increasing
in some region. This allows to define the boundaries of structural
anomaly at given temperature as minimum and maximum of excess
entropy.

The regions where these anomalies take place form nested domains
in the density-temperature \cite{deben2001} (or
pressure-temperature \cite{netz}) planes: the density anomaly
region is inside the diffusion anomaly domain, and both of these
anomalous regions are inside a broader structurally anomalous
region. It is reasonable to relate this kind of behavior to the
orientational anisotropy of the potentials, however, a number of
studies demonstrate water-like anomalies in fluids that interact
through spherically symmetric potentials
\cite{8,9,10,11,12,13,14,15,16,17,18,19,20,bar2,21,22,23,24,25,26,
barb2008-1,barb2008,yan2005,buld2009,FFGRS2008,RS2002,RS2003,FRT2006,GFFR2009,fr1,fr2}.

As it was discussed in many works (see, for example, the reviews
\cite{fr1} and \cite{buld2009}) the presence of two length scales
in the core-softening potential may be the origin of water-like
anomalies: a larger one, associated with the external finite
repulsion (effective at lower pressures and temperatures), and a
smaller one, related to the particle hard core (dominant at higher
pressures and temperatures). In those thermodynamic regimes where
the two length scales are both partially effective and thus are
competing with each other, a system of particles interacting
through such potentials behaves, in many respects, as a mixture of
two species of different sizes. This leads to the existence of two
competing local structures: an expanded structure characterized by
large open spaces between particles, and a collapsed structure in
which particles are spaced more closely. The evolution of these
structures under changing the thermodynamic conditions can result
in the anomalous behavior. For example, as it was shown in many
works (see, for example, \cite{bagchi2}), the low temperature
thermodynamic anomalies of liquid water arises from the
intermittent fluctuation between its high density and low density
forms, consisting largely of 5-coordinated and 4-coordinated water
molecules, respectively.

However, it should be noted that in general the existence of two
length scales is not enough to mark the occurrence of the
anomalies. For example, for the models studied in
Ref.~\cite{prest} it was shown that the existence of two distinct
repulsive length scales is not a necessary condition for the
occurrence of anomalous phase behavior.

The problem of anomalous behavior of core-softened fluids was
widely discussed in literature (see, for example, the recent
review \cite{fr1}). It was shown that for some systems the
anomalies take place while for others do not \cite{fr1}. In this
respect the question of criteria of anomalous behavior appearance
remains the central one. However, another important point is still
lacking in the literature - the behavior of anomalies along
different thermodynamic trajectories. Here we call as "trajectory"
a set of points belonging to some path in $(P,\rho,T)$ space. For
example, the set of points belonging to the same isotherm we call
as "isothermal trajectory" or shortly isotherm.

In our previous work we showed \cite{FR2011,werostr} that
anomalies can exist along some trajectories while along others the
liquid behaves as a simple one. Taking into account this result,
it is interesting to study the behavior of the quantities
demonstrating anomalies along the different physically significant
trajectories (isotherms, isochors, isobars and adiabats). This
investigation will allow to get deeper understanding of the
relations between anomalous behavior and thermodynamic parameters
of the system which spread light on the connection between
thermodynamic, structural and dynamic properties of liquids.

\section{II. System and Methods}

The simplest form of core-softened potential is the so called
Repulsive Step Potential which is defined as following:
\begin{equation}
U(r)=\left\{
\begin{array}{lll}
\infty , & r\leq d \\
\varepsilon , & d <r\leq \sigma  \\
0, & r>\sigma%
\end{array}%
\right.  \label{1}
\end{equation}
where $d$ is the diameter of the hard core, $\sigma$ is the width
of the repulsive step,  and  $\varepsilon$ is its height. In the
low-temperature limit $\tilde{T}\equiv k_BT/\varepsilon<<1$  the
system reduces to a hard-sphere system with hard-sphere diameter
$\sigma$, whilst in the limit $\tilde{T}>>1$ the system reduces to
a hard-sphere model with a hard-sphere diameter $d$. For this
reason, melting at high and low temperatures follows simply from
the hard-sphere melting curve $P=cT/\sigma'^3$, where $c \approx
12$ and $\sigma'$ is the relevant hard-sphere diameter ($\sigma$
and $d$, respectively). A changeover from the low-$T$ to high-$T$
melting behavior should occur for $\tilde{T} ={\mathcal O}(1)$.
The precise form of the phase diagram depends on the ratio
$s\equiv \sigma/d$. For large enough values of $s$ one should
expect to observe in the resulting melting curve a maximum that
should disappear as $s\rightarrow 1$. The phase behavior in the
crossover region may be very complex, as  shown
in~\cite{FFGRS2008,GFFR2009}.

In the present simulations we have used a smoothed version of the
repulsive step potential (Eq.~(\ref{1})), which has the form:
\begin{equation}
U(r)=
\varepsilon\left(\frac{d}{r}\right)^{n}+\frac{1}{2}\varepsilon\left(1-\tanh\left(k_0
\left(r-\sigma_s \right)\right)\right), \label{2}
\end{equation}
where $n=14,k_0=10$. We have considered $\sigma_s= 1.35$. Here and
below we refer to this potential as to Smooth Repulsive Shoulder
System (SRSS).

In the remainder of this paper we use the dimensionless
quantities: $\tilde{{\bf r}}\equiv {\bf r}/d$, $\tilde{P}\equiv P
d^{3}/\varepsilon ,$ $\tilde{V}\equiv V/N d^{3}\equiv
1/\tilde{\rho}$. As we will only use these reduced variables, we
omit the tildes.

In Refs.~\cite{FFGRS2008,GFFR2009},  phase diagrams of SRSS models
were reported for $\sigma_s=1.15, 1.35, 1.55, 1.8$.

\begin{figure}
\includegraphics[width=7cm, height=7cm]{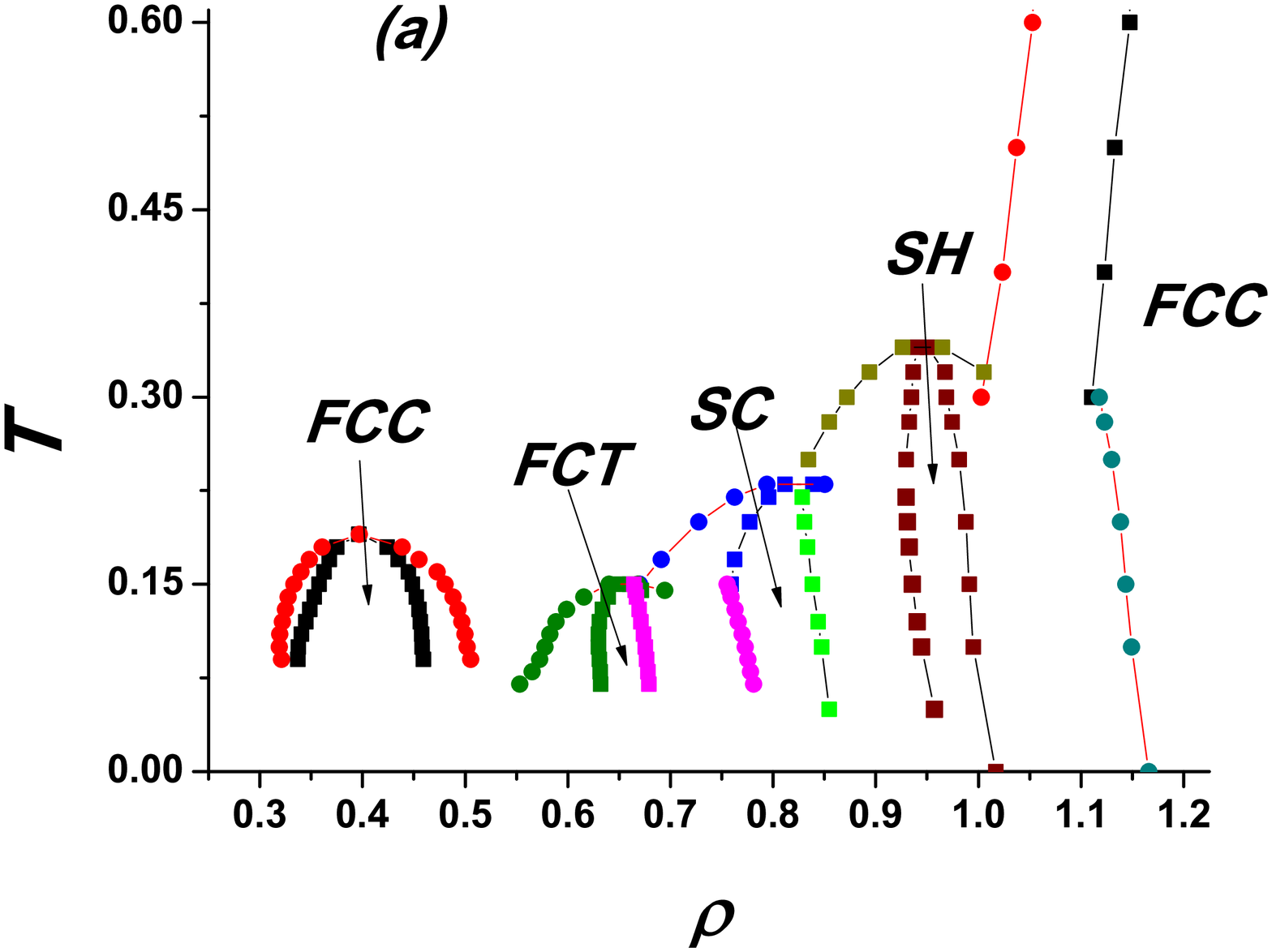}

\includegraphics[width=7cm, height=7cm]{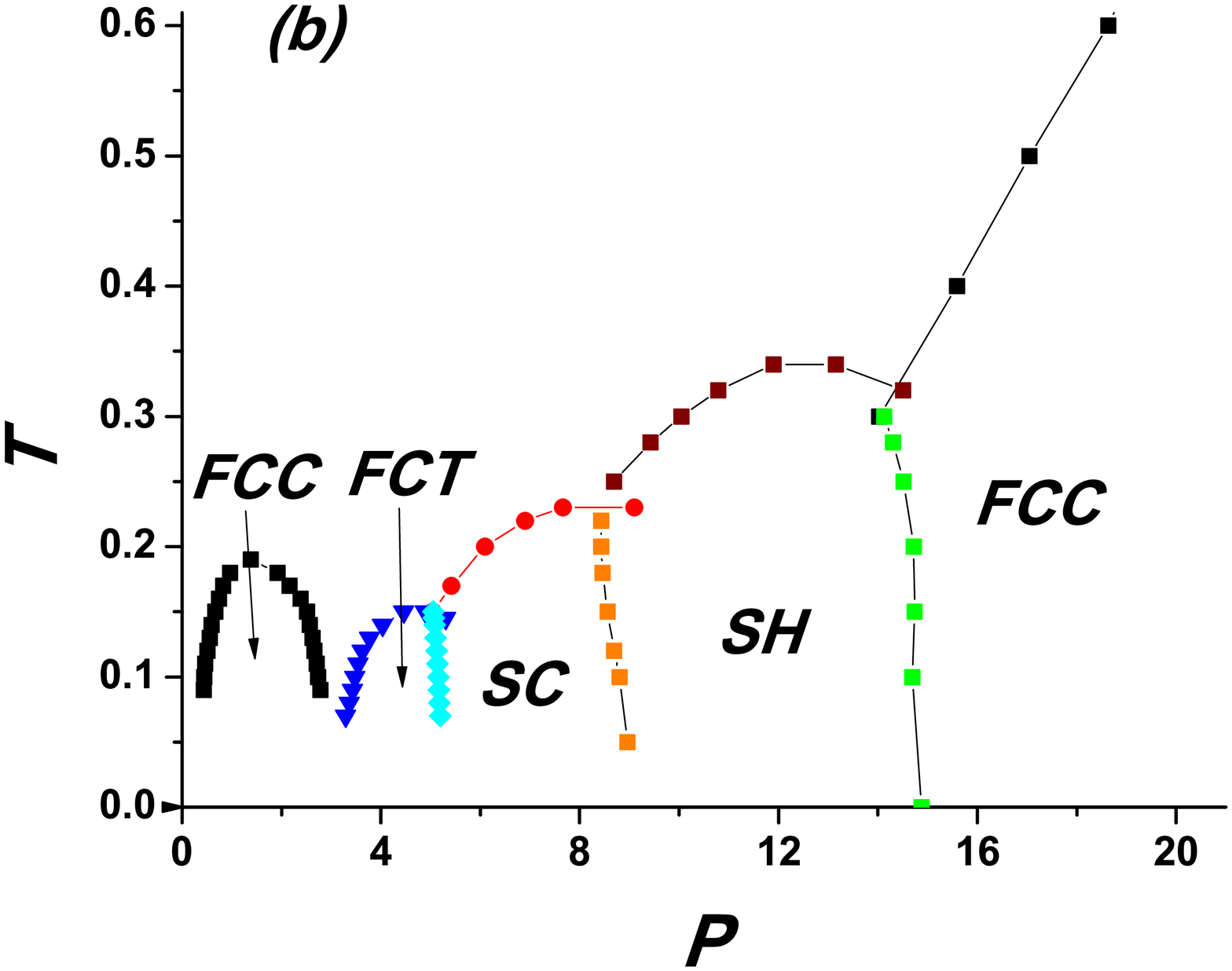}%

\caption{\label{fig:fig1a} (Color online). Phase diagram of the
system of particles interacting through the potential (\ref{2})
with $\sigma_s= 1.35$ in $\rho-T$ (a) and $P-T$ (b) planes.}
\end{figure}

Fig.~\ref{fig:fig1a} shows the phase diagrams that we obtain from
the free-energy calculations for $\sigma_s=1.35$. In fact, the
phase diagrams for $\sigma_s=1.15, 1.35, 1.55, 1.8$ were already
reported in Refs.~\cite{FFGRS2008,GFFR2009}. We show these phase
diagrams here too because they provide the ``landscape'' in which
possible ``water'' anomalies should be positioned.

Fig.~\ref{fig:fig1a}(a) shows the phase diagram of the system with
$\sigma_s=1.35$ in the $\rho-T$ plane. There is a clear maximum in
the melting curve at low densities. The phase diagram consists of
two isostructural FCC domains corresponding to close packing of
the small and large spheres separated by a sequence of structural
phase transitions. This phase diagram was discussed in detail in
our previous publications \cite{FFGRS2008,GFFR2009}. It is
important to note that there is a region of the phase diagram
where we have not found any stable crystal phase. The results of
Ref.~ \cite{FFGRS2008} suggest that a glass transition occurs in
this region with vitrification temperature $T_g=0.079$ at
$\rho=0.53$. The apparent glass-transition temperature is above
the melting point of the low-density FCC and FCT phases. If,
indeed, no other crystalline phases are stable in this region, the
``glassy'' phase that we observe would be thermodynamically
stable. This is rather unusual for one-component liquids. In
simulations, glassy behavior is usually observed in metastable
mixtures, where crystal nucleation is kinetically suppressed. One
could argue that, in the glassy region, the present system behaves
like a ``quasi-binary'' mixture of spheres with diameters $d$ and
$\sigma_s$ and that the freezing-point depression is analogous to
that expected in a binary system with an eutectic point: there are
some values of the diameter ratio such that crystalline structures
are strongly unfavorable and the glass phase could then be stable
even at very low temperatures. The glassy behavior in the
reentrant liquid disappears at higher temperatures.

In the present study a system of particles interacting via the
potential with "hard" core, repulsive shoulder and attractive well
is also investigated. This potential represents a generalization
of our previous SRSS model \cite{FFGRS2008,GFFR2009} and we call
it Smooth Repulsive Shoulder System with Attractive Well (SRSS-AW)
potential \cite{FRT2011}.

The general form of the potential is written as
\begin{eqnarray}
U(r)&=&\varepsilon\left(\frac{\sigma}{r}\right)^{14}+\varepsilon\left(\lambda_0-
\lambda_1\tanh(k_1\{r-\sigma_1\})\right.+\nonumber\\
&+&\left.\lambda_2 \tanh(k_2\{r-\sigma_2\})\right). \label{22}
\end{eqnarray}

We consider only the potentials with $\sigma_1=1.35$ (see Table
1).

Before to proceed, let us consider the analytic condition for core
softening proposed in Ref.~\cite{deben1991}:
\begin{equation}
r U''(r)+U'(r)<0, r_2>r>r_1 \label{deben1}
\end{equation}
and also
\begin{equation}
U''(r)>0 \label{deben2}
\end{equation}
for $r<r_1$ and $r>r_2$, where $r_1$ and $r_2$ denote two
repulsive length scales. In the systems satisfying these
conditions there are two local structures which compete with each
other. In this case the system behaves as the mixture of two types
of particles with effective radii $r_1$ and $r_2$ (see, for
example, Ref.~\cite{FFGRS2008}), and one can expect to find a
reentering melting and other anomalous behaviors.

It may be easily seen that for the potentials (1-3) (see Table 1)
the conditions (\ref{deben1}) and (\ref{deben2}) are satisfied for
$\sigma_1\geq 1.16$, so in the case $\sigma_1=1.35$ the anomalies
do exist \cite{GFFR2009,FRT2011}.

However, as it was mentioned above, in general the conditions
(\ref{deben1}) and (\ref{deben2}) are not enough to mark the
occurrence of the anomalies \cite{prest}.

In Refs.~\cite{fr1,fr2} the extensive study of the softness
dependence of the anomalies for the continuous shouldered well
potential was presented and different criteria for the appearance
of the anomalies was analyzed. It was shown that for the more
steeper soft-core the regions of the density and diffusion
anomalies become more narrow, while the region of the structural
anomaly is only weekly affected, and  the disappearance of the
density and diffusion anomalies for the steeper potentials is due
to a more structured short-range order. At the same time, it may
be shown that the conditions (\ref{deben1}) and (\ref{deben2}) are
satisfied for the potentials considered in Refs.~\cite{fr1,fr2}.
For the steeper potentials the range between $r_1$ and $r_2$ is
more narrow, and the effective diameter of the soft core is
larger. However, as it was shown earlier \cite{GFFR2009}, the
increasing of the soft core diameter leads to the disappearance of
the anomalies in the range of the thermodynamic stability of the
system. It seems that the search of the adequate criterium
relating the appearance of the anomalies with the intermolecular
potential is very interesting and important problem.

\begin{figure}
\includegraphics[width=8cm]{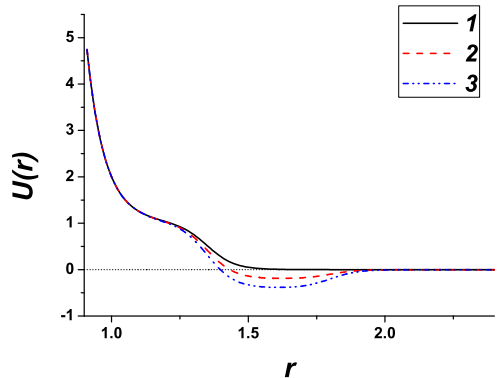}%

\caption{\label{fig:fig22} (Color online) Family of the potentials
with $\sigma_1=1.35$ and different attractive wells. The curves
are numerated in accordance with Table 1.}
\end{figure}

We consider the system with the step $\sigma_1=1.35$. The
parameters $k_1=k_2=10.0$ are fixed while parameters $\sigma_2,
\lambda_0$, $\lambda_1$ and $\lambda_2$ are varied to get the
different potential shape. Three sets of parameters are
considered. They are summarized in Table 1. Fig.~\ref{fig:fig22}
shows the potentials for the step width $\sigma_1=1.35$
respectively. The parameters are chosen in such a way that the
depth of attractive well becomes larger (see Table 1 and
Fig.~\ref{fig:fig22}). Below we denote the systems with different
parameters as system 1, system 2 and so on in accordance with the
Table 1. The choice of the potential parameters was dictated only
by the convenience for the analysis of the qualitative influence
of the attraction on the properties of the system.

\begin{table}
\begin{tabular}{|c|c|c|c|c|c|c|}
  \hline
  number & $\sigma_1$ & $\sigma_2$& $\lambda_0$  & $\lambda_1$ & $\lambda_2$ & well depth \\
  \hline
  1 & 1.35 & 0 & 0.5 & 0.5 & 0 & 0\\
  2 & 1.35 & 1.80 & 0.5 & 0.60 & 0.10 & 0.20\\
  3 & 1.35 & 1.80 & 0.5 & 0.7 & 0.20 & 0.4\\
  \hline

\end{tabular}

\caption{The potential parameters used in simulations (Eq.
(\ref{22})).}

\end{table}

\begin{figure}
\includegraphics[width=8cm]{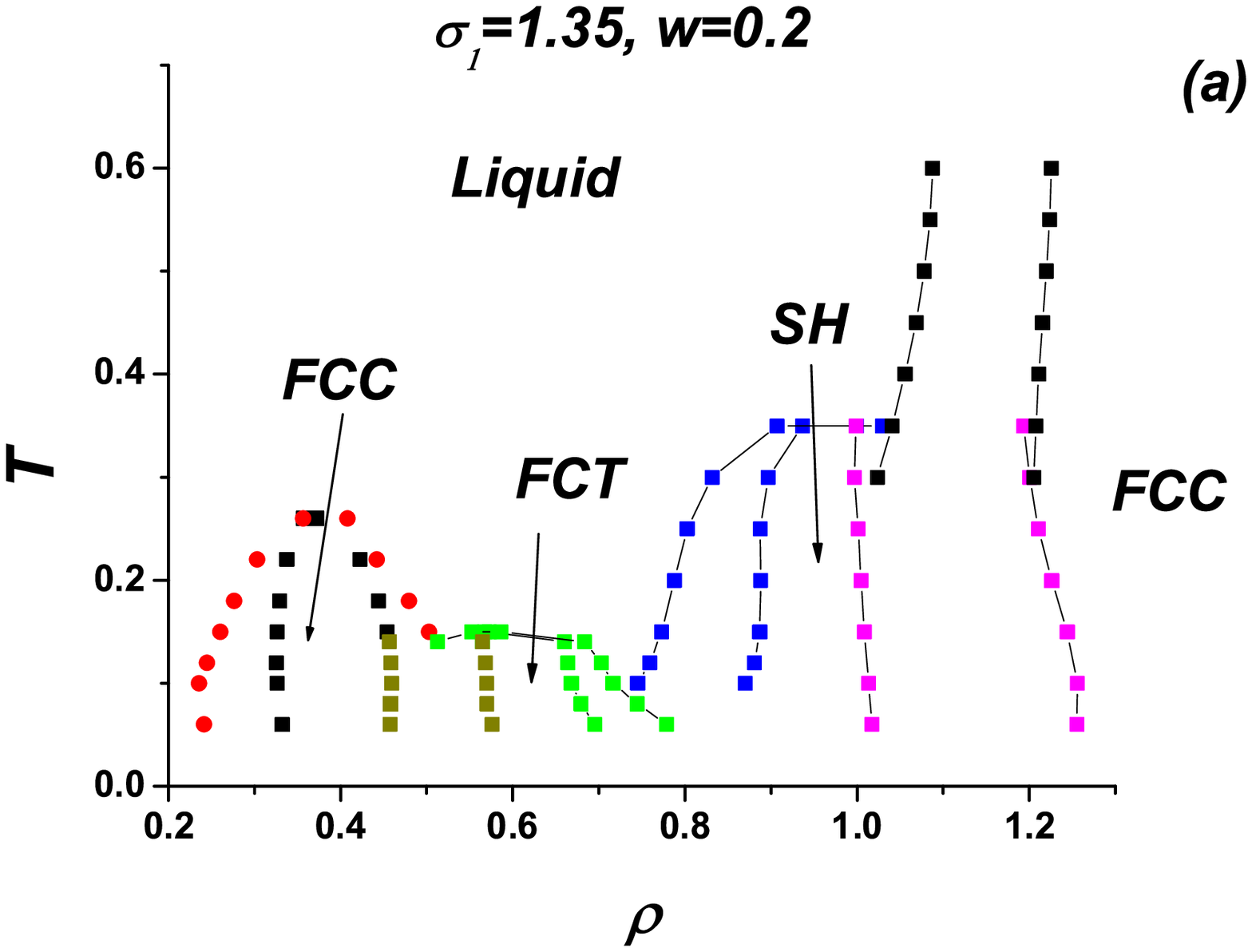}%

\includegraphics[width=8cm]{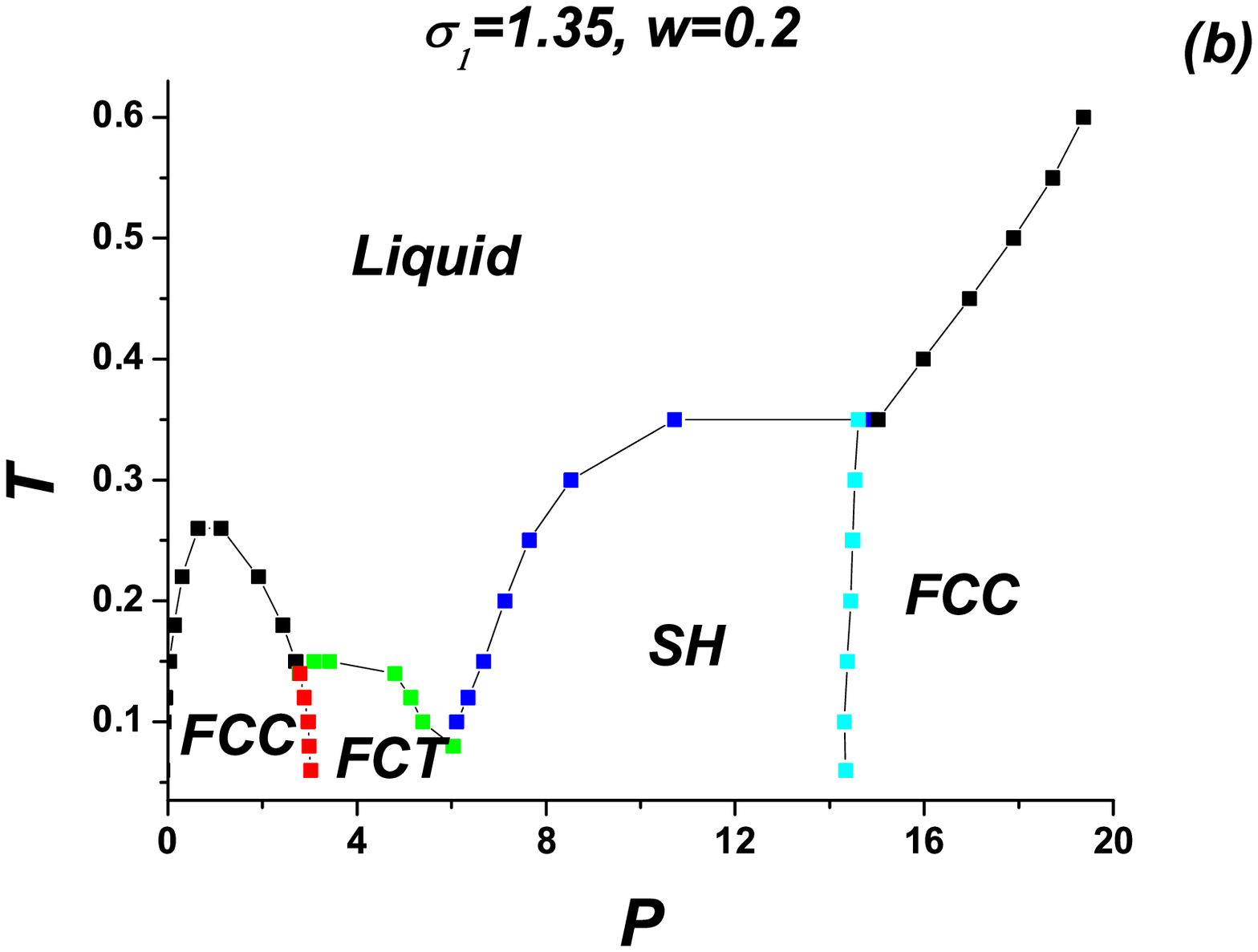}%

\caption{\label{fig:fig2a} (Color online) Phase diagram of the
system 2 (Table 1) in (a) $\rho - T$ and (b) $P - T$ coordinates.}
\end{figure}

\begin{figure}
\includegraphics[width=8cm]{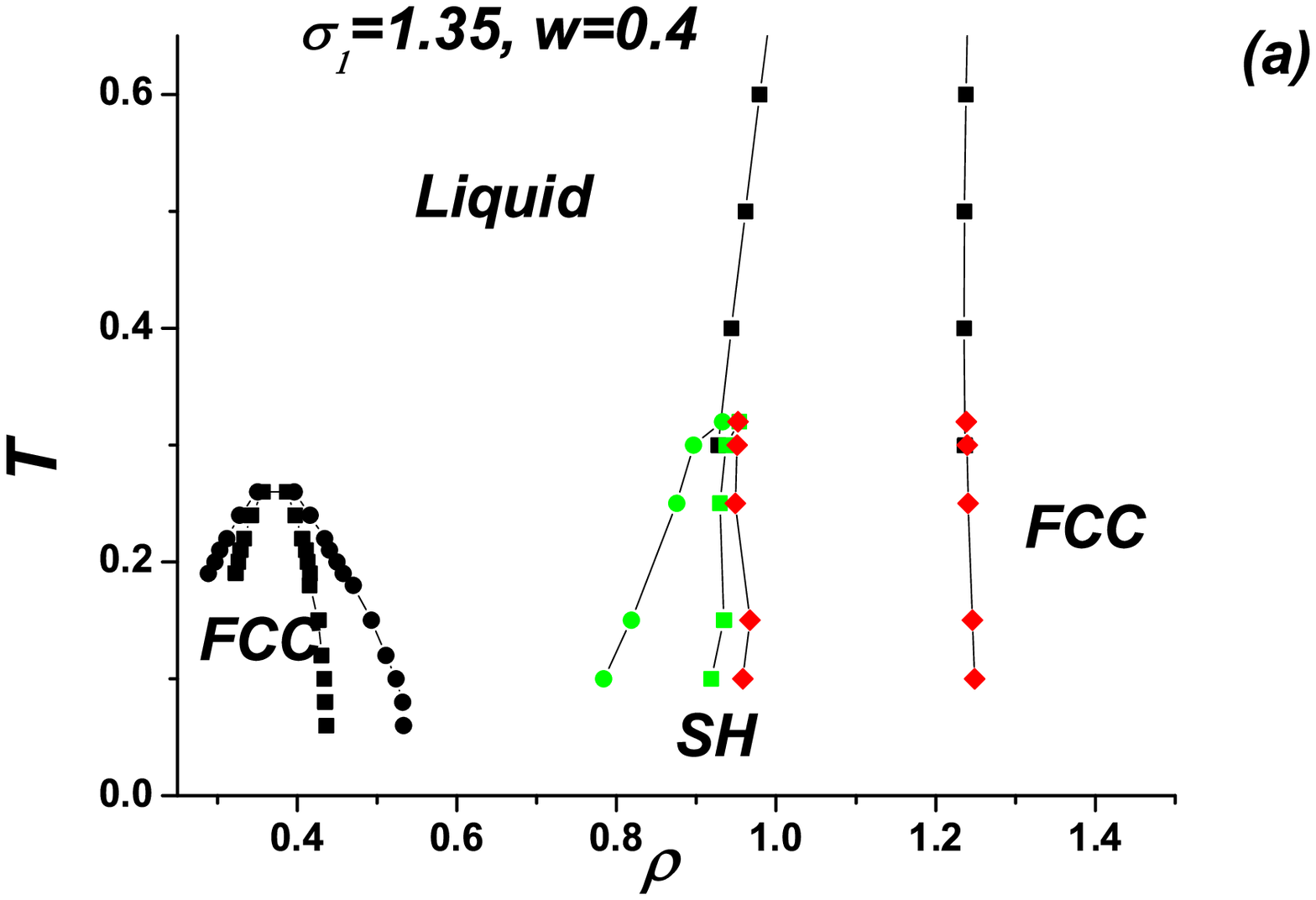}%

\includegraphics[width=8cm]{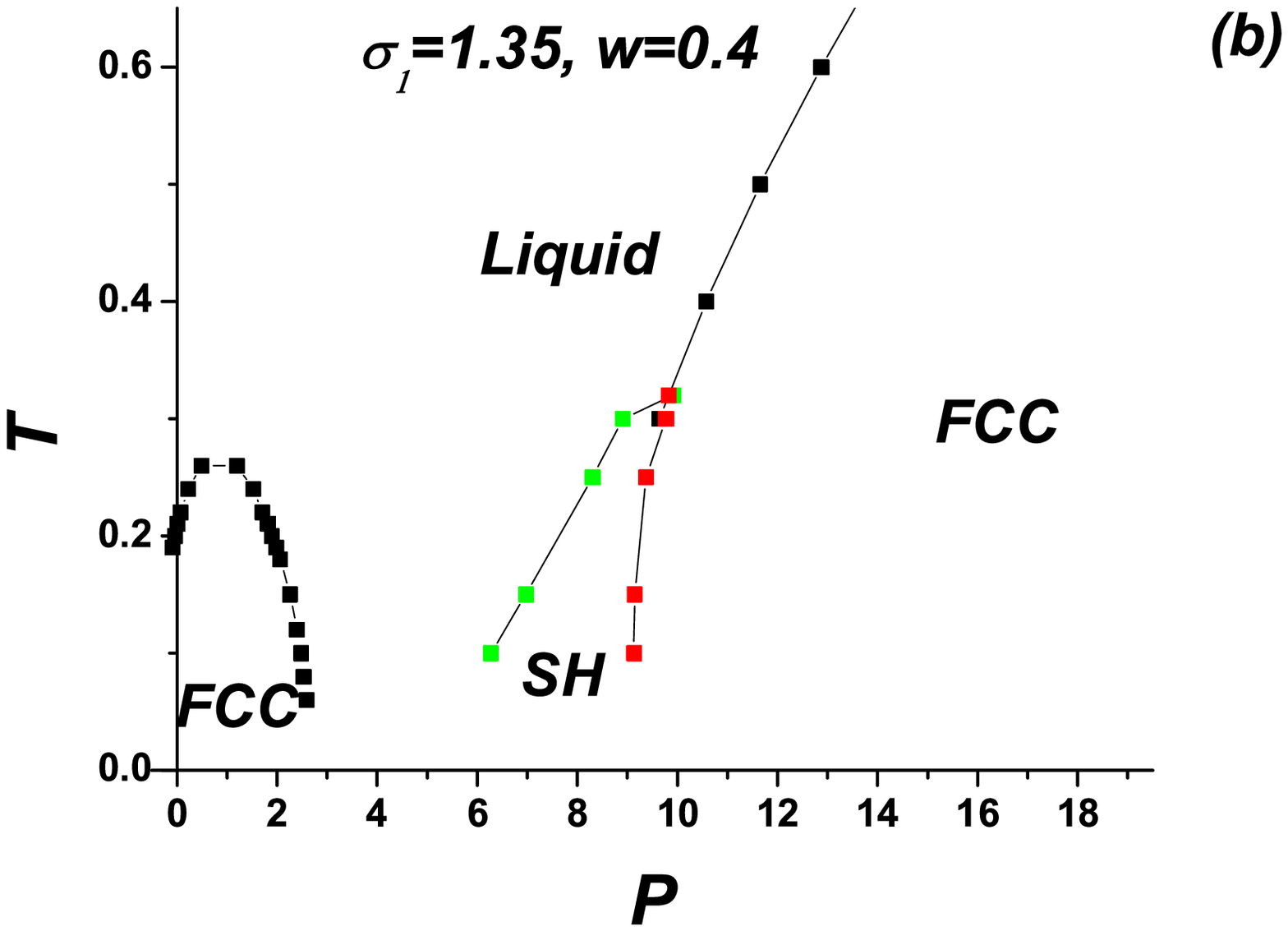}%

\caption{\label{fig:fig3a} (Color online) Phase diagram of the
system 3 (Table 1) in (a) $\rho - T$ and (b) $P - T$ coordinates.}
\end{figure}

In our previous publications \cite{FFGRS2008,GFFR2009} we
discussed the phase diagrams of several purely repulsive systems,
i.e. the systems with zero well depth. The complexity of these
phase diagrams was shown. The systems with attractive well were
considered in \cite{FRT2011}. For the completeness we present the
evolution of the system behavior with increasing attraction in
Figs.~\ref{fig:fig2a} and \ref{fig:fig3a}.

The transition lines were determined from the free energy
calculations \cite{book_fs,fladd}.

We simulate the system in $NVT$ ensemble using Monte-Carlo method.
The number of particles in the liquid or gas state simulation was
set to $500$ or $1000$ and for crystal phases it varied between
$500$ and $1000$ depending on the structure. The system was
equilibrated for $10^6$ MC step and the data were collected during
$10^5$ MC steps.

In order to find the transition points we carry out the free
energy calculations for different phases and construct a common
tangent to them. For the purely repulsive potentials we computed
the free energy of the liquid by integrating the equation of state
along an isotherm \cite{book_fs,fladd}:
$\frac{F(\rho)-F_{id}(\rho)}{Nk_BT}=\frac{1}{k_BT}\int_{0}^{\rho}\frac{P(\rho')-\rho'
k_BT}{\rho'^2}d\rho'$. In the case of potentials which contain an
attractive part the situation is more complicated because of the
possible gas - liquid transition. In order to avoid the
difficulties connected to this transition we carry out calculation
of free energies at high temperature above the gas - liquid
critical point and then calculate the free energies by integrating
the internal energies along an isochor \cite{book_fs,fladd}:
$\frac{F(T_2)-F(T_1)}{k_BT}=\int_{T_1}^{T_2}U(T,N,V)d(\frac{1}{T})$.

Free energies of different crystal phases were determined by the
method of coupling to the Einstein crystal \cite{book_fs,fladd}.

To improve the statistics (and to check for internal consistency)
the free energy of the solid was computed at many dozens of
different state points and fitted to multinomial function. The
fitting function we used is $a_{p,q}T^pV^q$, where $T$ and
$V=1/\rho$ are the temperature and specific volume and powers $p$
and $q$ are related through $p+q \leq N$. The value $N$ we used
for the most of calculations is $5$.

The diffusion anomaly is also discussed in the article. Since the
diffusion coefficient can not be measured in Monte Carlo
simulations, molecular dynamics is applied. The core-softened
systems are characterized by a complex energy landscape. This
makes difficult to simulate the system at low temperatures. In
order to avoid this problem, parallel tempering technique is used
\cite{book_fs,fladd}. The diffusion is measured along a set of
isochors between the densities $\rho=0.3$ and $\rho=0.8$. The
temperatures used are confined between $T=0.15$ and $T=0.8$.

We simulate a system of $864$ particles in a cubic box. Each
parallel tempering run consists of $16$ exchanges between $8$
different temperatures. Between the exchanges the system evolves
for $4 \cdot 10^6$ steps. The first $3 \cdot 10^6$ steps are used
for equilibration. The time step is $dt=0.0005$. In order to keep
the temperature constant Andersen thermostat is used during the
equilibration. Summing up all simulations done and taking into
account exchange of temperatures in parallel tempering runs, we
collect more then a hundred measurements along each isochor which
gives good statistics. The diffusion coefficient along isochors is
approximated by a $9-$th order polynomial of temperature. Then the
data are rearranged along isotherms.

\section{III. Results and discussion}

\subsection{Diffusion anomaly}

The diffusion anomaly of the SRSS was discussed in several our
previous articles. In the Refs. \cite{FFGRS2008,GFFR2009,FRT2011}
we showed that the diffusion anomaly takes place at the shoulder
width $\sigma_1=1.35$ while in the work \cite{weros} the breakdown
of the Rosenfeld scaling for this system was demonstrated. Further
discussion of the Rosenfeld scaling was reported in the works
\cite{FR2011,werostr}. The main idea of these papers is that the
anomalous diffusion behavior present along low-temperature
isotherms while along isochors diffusion coefficient is
monotonous. As a result, Rosenfeld scaling is valid along isochors
and high-temperature isotherms, however, it breaks down for the
isotherms with anomalies. It means that appearance or not of
anomalies at some trajectory can cause physically different
behavior of the system. As a result, following different
trajectories, we can or can not observe some effects. It is of
particular importance since experimental works deal mostly with
isobars and isotherms while theoretical studies - with isochors
and isotherms. Taking this into account, one can expect that some
of the effects observed along isobars are not visible along
isotherms and isochors which makes us confused while comparing
experimental results with theoretical predictions. Here we extend
the study of anomalies to four different physically meaningful
trajectories: isotherms, isochors, isobars and adiabats and
consider both SRSS and SRSS-AW potentials.

\begin{figure}
\includegraphics[width=7cm, height=7cm]{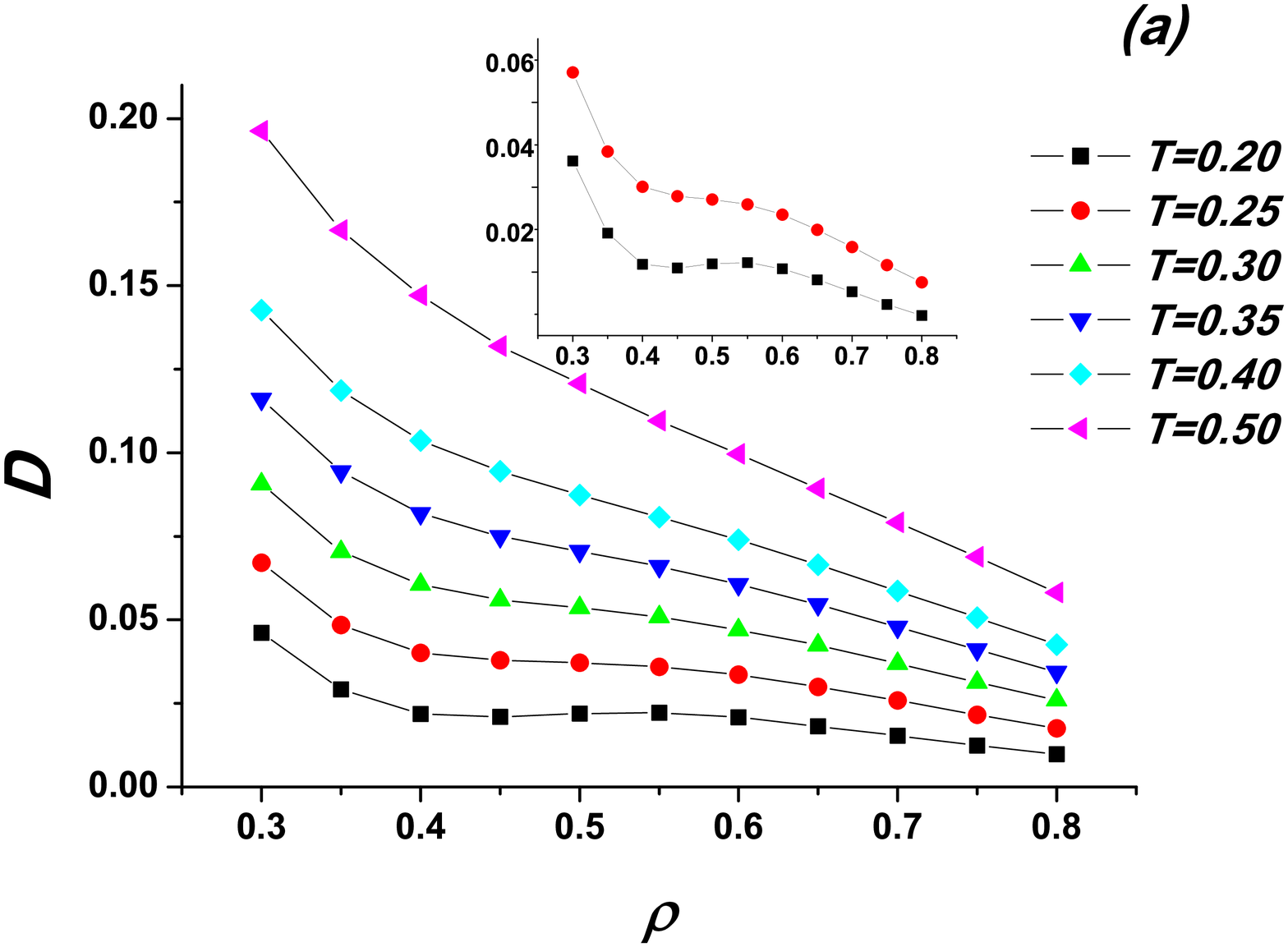}%

\includegraphics[width=7cm, height=7cm]{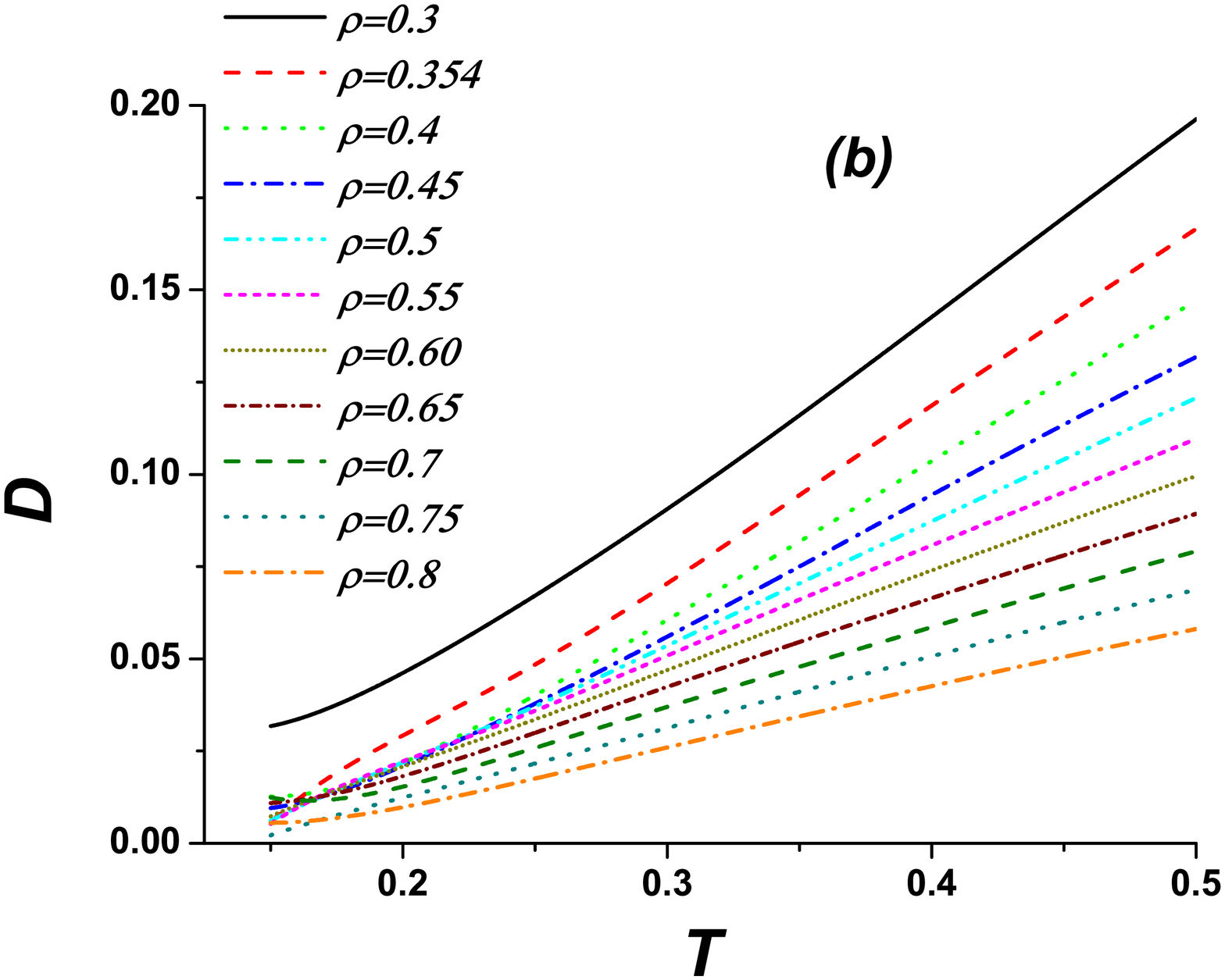}%

\includegraphics[width=7cm, height=7cm]{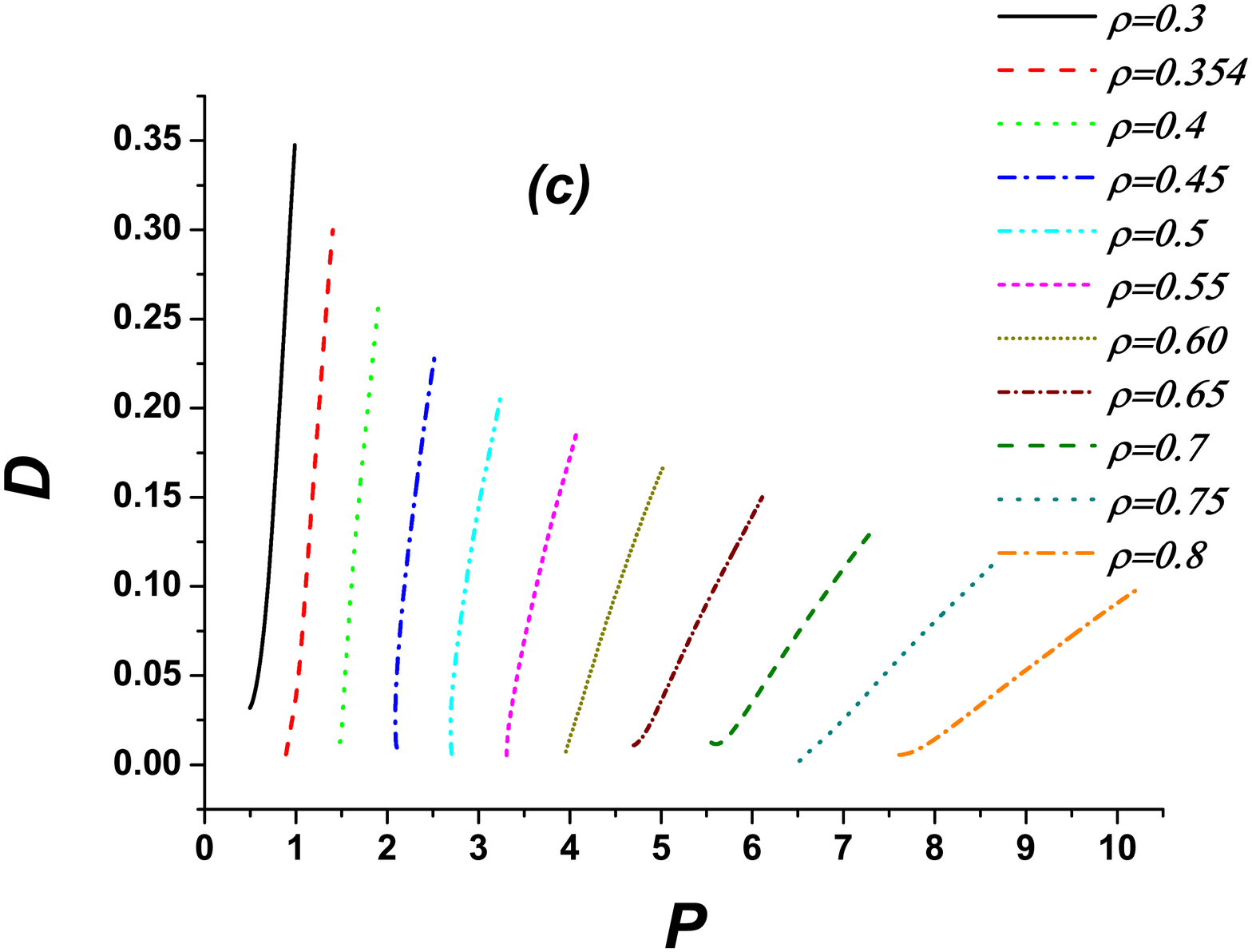}%

\caption{\label{fig:fig1} (Color online). Diffusion coefficient of
the SRSS system along (a) isotherms and (b) and (c) - isochors.
The insert in (a) shows the low temperature isotherms.}
\end{figure}

\begin{figure}
\includegraphics[width=7cm, height=7cm]{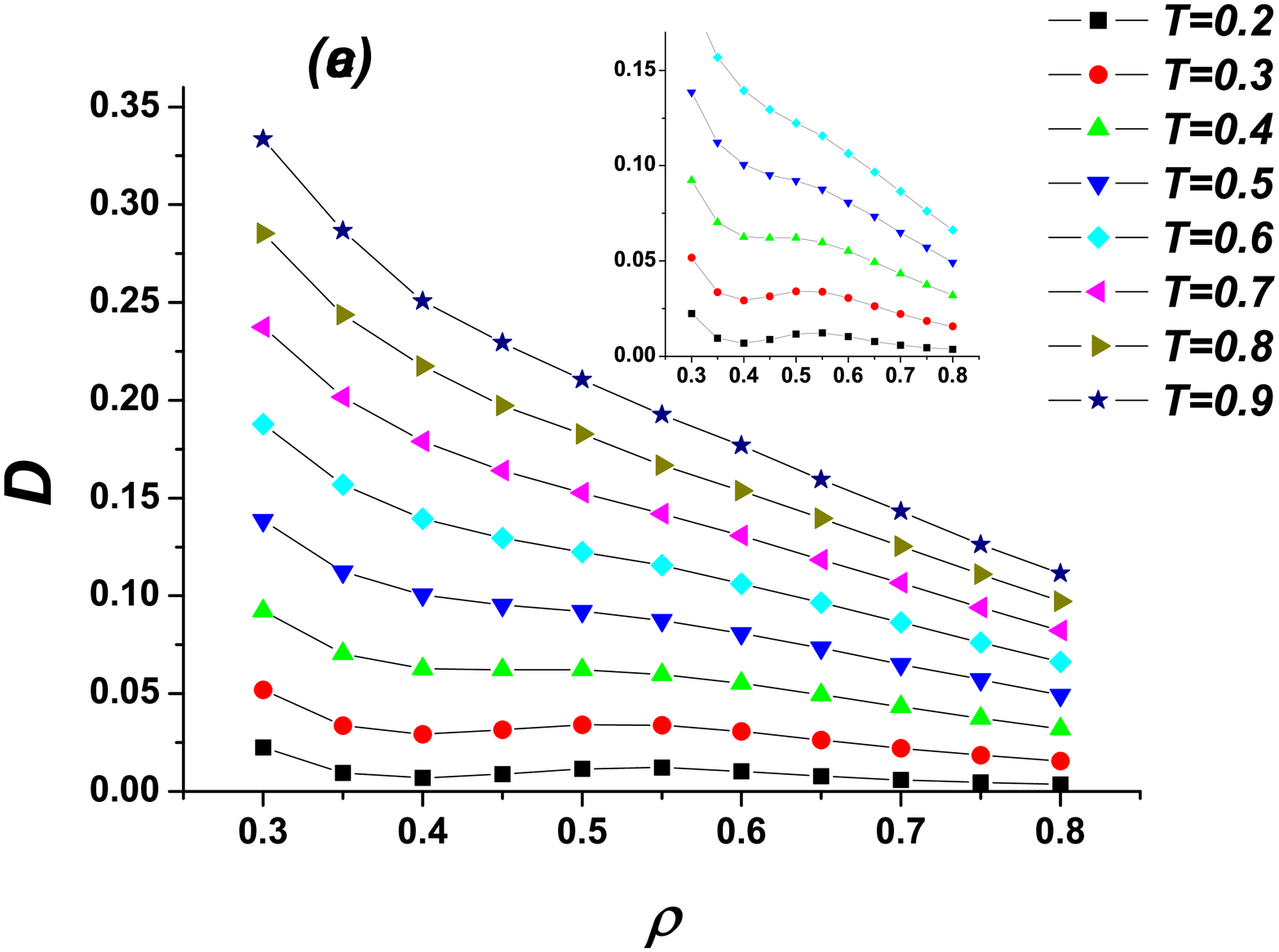}%

\includegraphics[width=7cm, height=7cm]{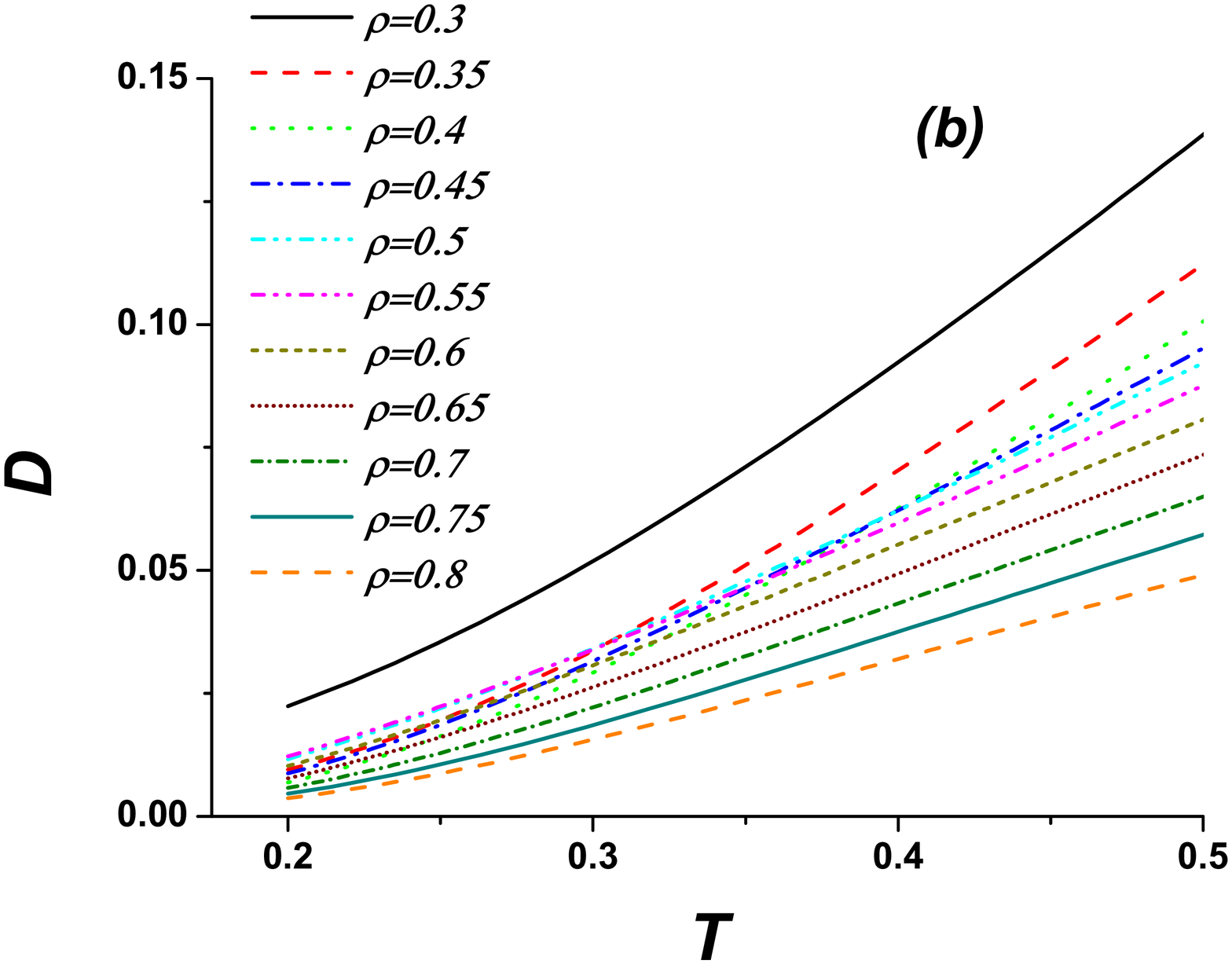}%

\includegraphics[width=7cm, height=7cm]{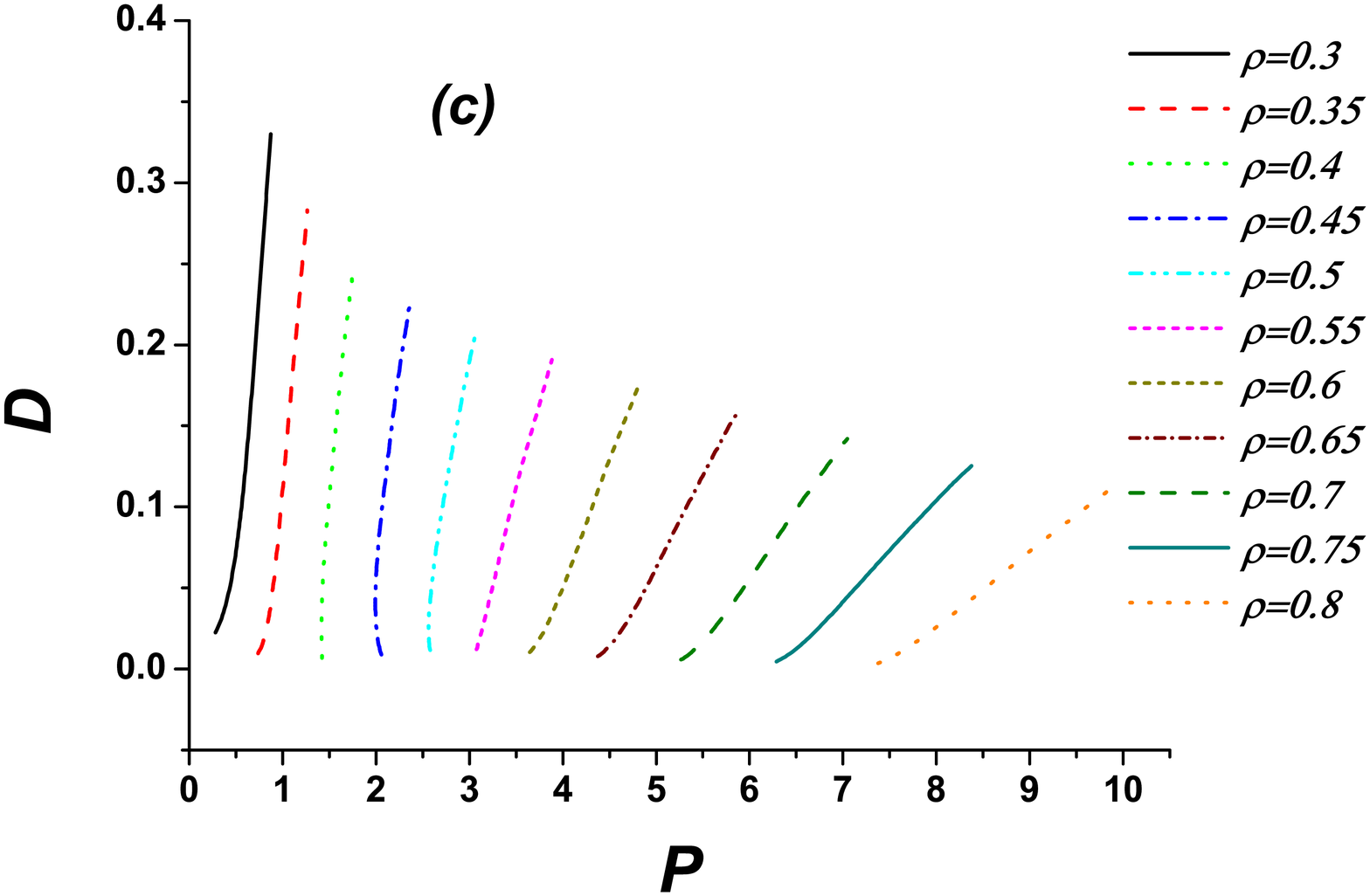}%

\caption{\label{fig:fig1at} (Color online). Diffusion coefficient
of the SRSS-AW system (system 3 in Table 1) along (a) isotherms
and (b) and (c) - isochors.}
\end{figure}

Fig.~\ref{fig:fig1}(a) shows the diffusion coefficient of SRSS
system along a set of isotherms for $\sigma_1=1.35$ (potential 1
in Table 1). One can see that the diffusion coefficient
demonstrates anomalous behavior for the temperatures below
$T=0.25$. On the other hand, if we look at the
Fig.~\ref{fig:fig1}(b) where the same diffusion coefficient data
are arranged along isochors as a function of temperature we do not
observe anomalies - the diffusion is a monotonous function of
temperature along isochors. However, some of the isochors cross.
One can see, that the cross of the isochors corresponds to the
densities between $\rho=0.4$ and $\rho=0.6$. Comparing it to the
isotherms we see that this is the region of anomalous diffusion.
It means that even if we do not see nonmonotonous behavior of
diffusion along isochors we can identify the presence of anomaly
from crossing of the isochors. However, this method seems to be
technically more difficult since we need to measure many points
belonging to different isochors rather then one isotherm. In
Fig.~\ref{fig:fig1}(c) the diffusion coefficient is shown as a
function of pressure along the isochores. One can see that in the
range of the densities between $\rho=0.4$ and $\rho=0.6$ the
slopes of the curves change sign. Isothermal and isochoric
behavior of diffusion in anomalous region for SRSS have already
been discussed in our previous publications \cite{FR2011,werostr}
and we give these plots here for the sake of completeness.

If attraction is added to the potential (systems 2 and 3 in Table
1), no new qualitative features are found, however, the anomalies
become more pronounced. One can see this, for example, in
Figs.~\ref{fig:fig1at} ((a), (b) and (c)), where the diffusion
coefficients of the SRSS-AW potential of the system 3 are shown
along the isotherms and isochores.

\begin{figure}
\includegraphics[width=7cm, height=7cm]{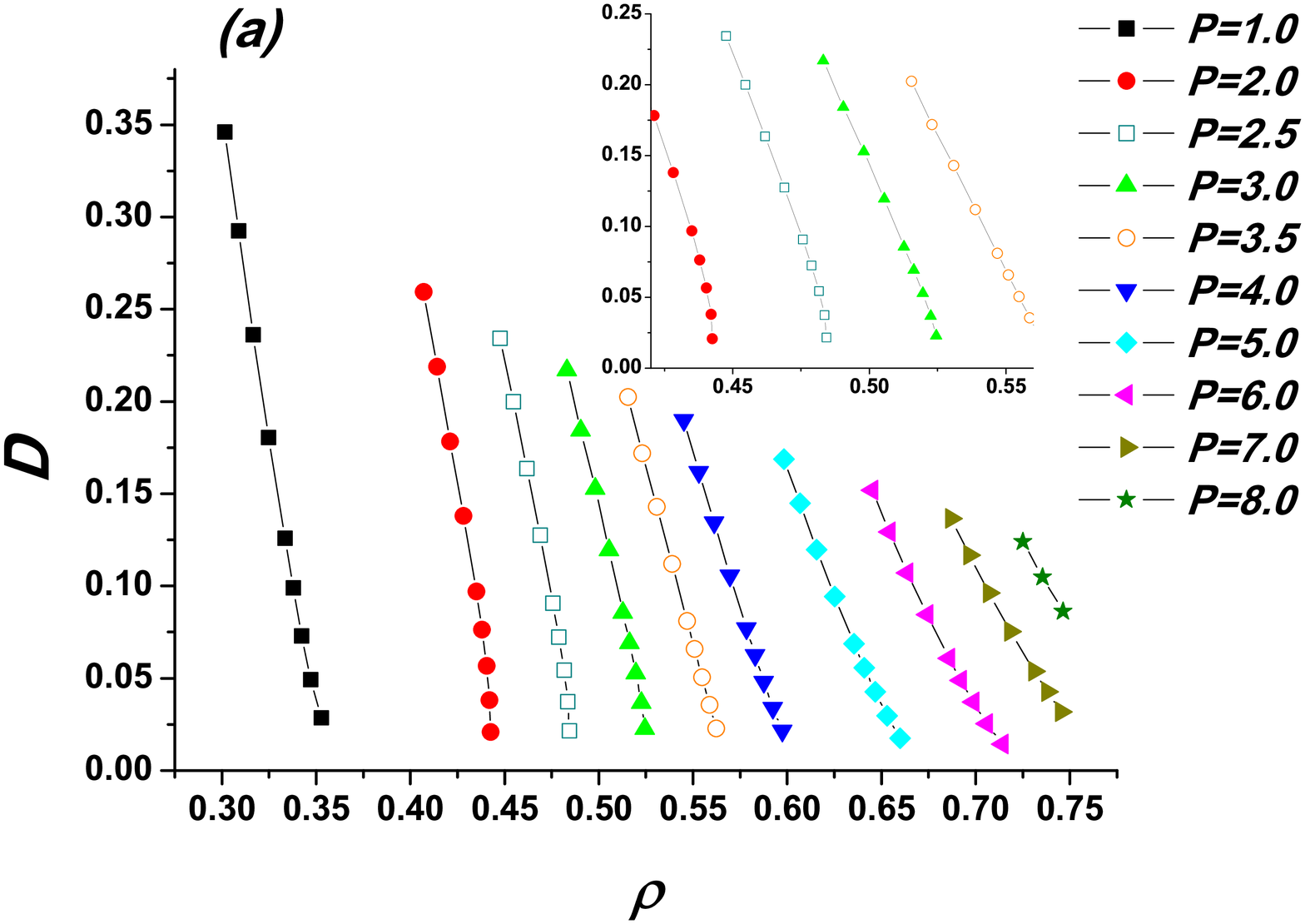}%

\includegraphics[width=7cm, height=7cm]{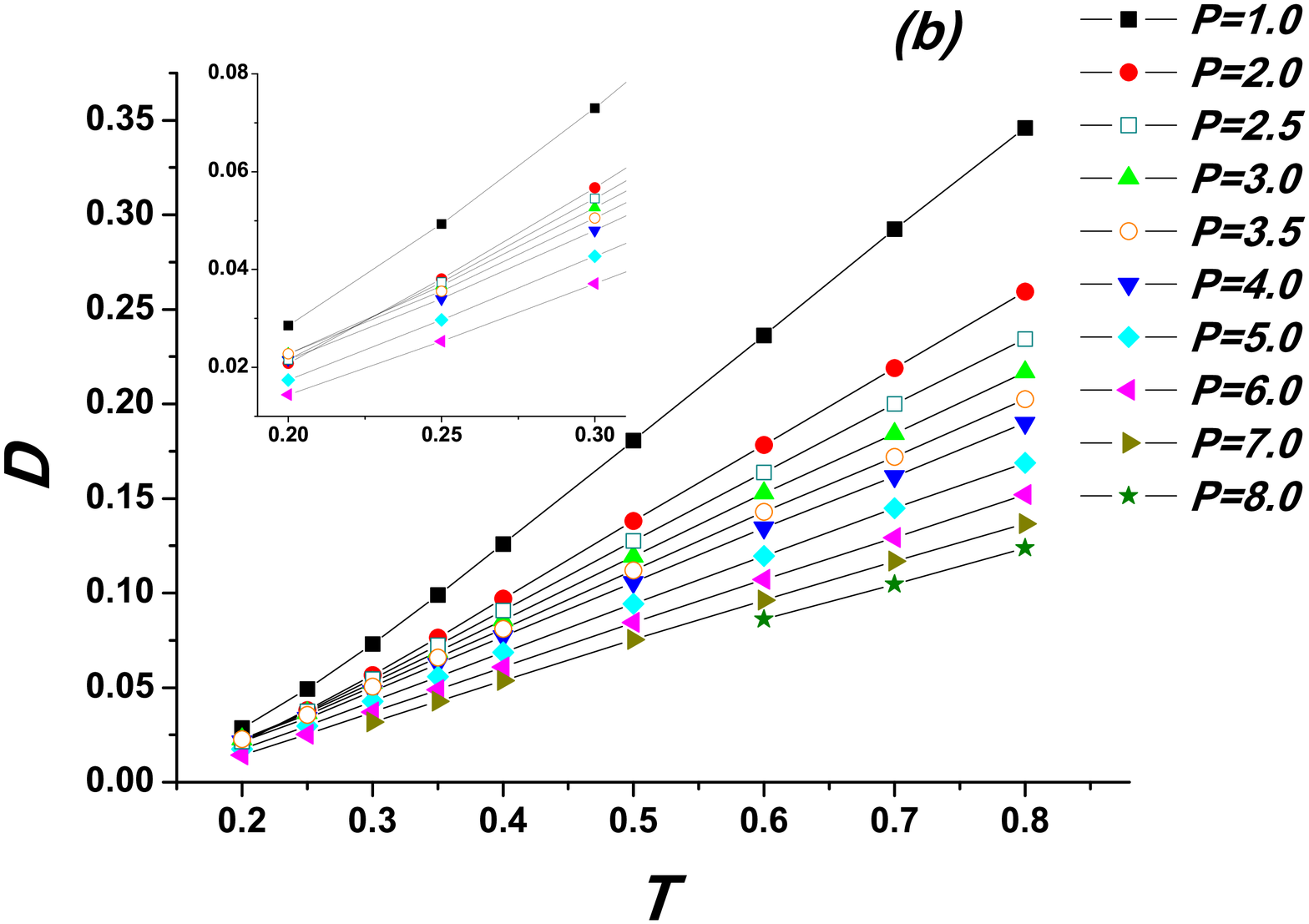}%
\caption{\label{fig:fig2} (Color online). Diffusion coefficient of
the SRSS system along isobars as a function of (a) density and (b)
temperature. The inserts show anomalous regions in the
corresponding coordinates.}
\end{figure}

Fig.~\ref{fig:fig2}(a) shows the diffusion coefficient along a set
of isobars as a function of density. The diffusion coefficient is
again monotonous. The slope of the curves is always negative.
However, as it can be seen from the insert, the slope approaches
infinity at low diffusions at pressures $P=2.0$ and $P=2.5$. This
corresponds to densities $0.45-0.50$ inside the anomalous region.
One can imagine that if we lower the temperatures along these
isobars we can observe change of the slope to positive one,
however, we do not have data for these temperatures.

Fig.~\ref{fig:fig2}(b) shows the diffusion coefficient along
isobars as a function of temperature. The situation is analogous
to the case of isochors: the curves are monotonous, however, they
intersect at low temperatures, corresponding to anomalous region
(see the insert in Fig.~\ref{fig:fig2}(b)). It means that if we
have the diffusion coefficient along isobars we can identify the
presence of anomalies by monitoring the intersections of the
curves. However, by the reasons discussed above this method is not
practically convenient.

\begin{figure}
\includegraphics[width=7cm, height=7cm]{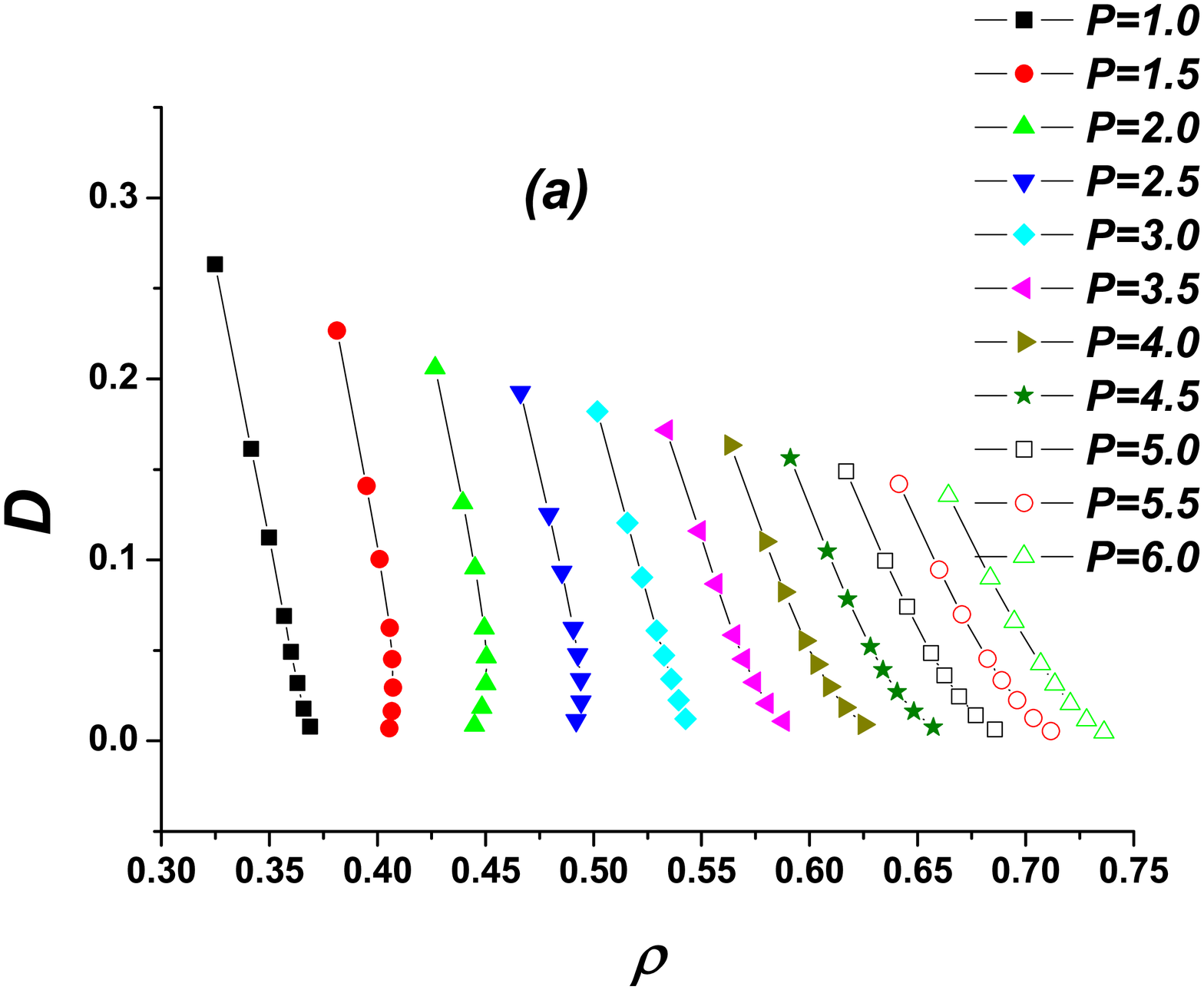}%

\includegraphics[width=7cm, height=7cm]{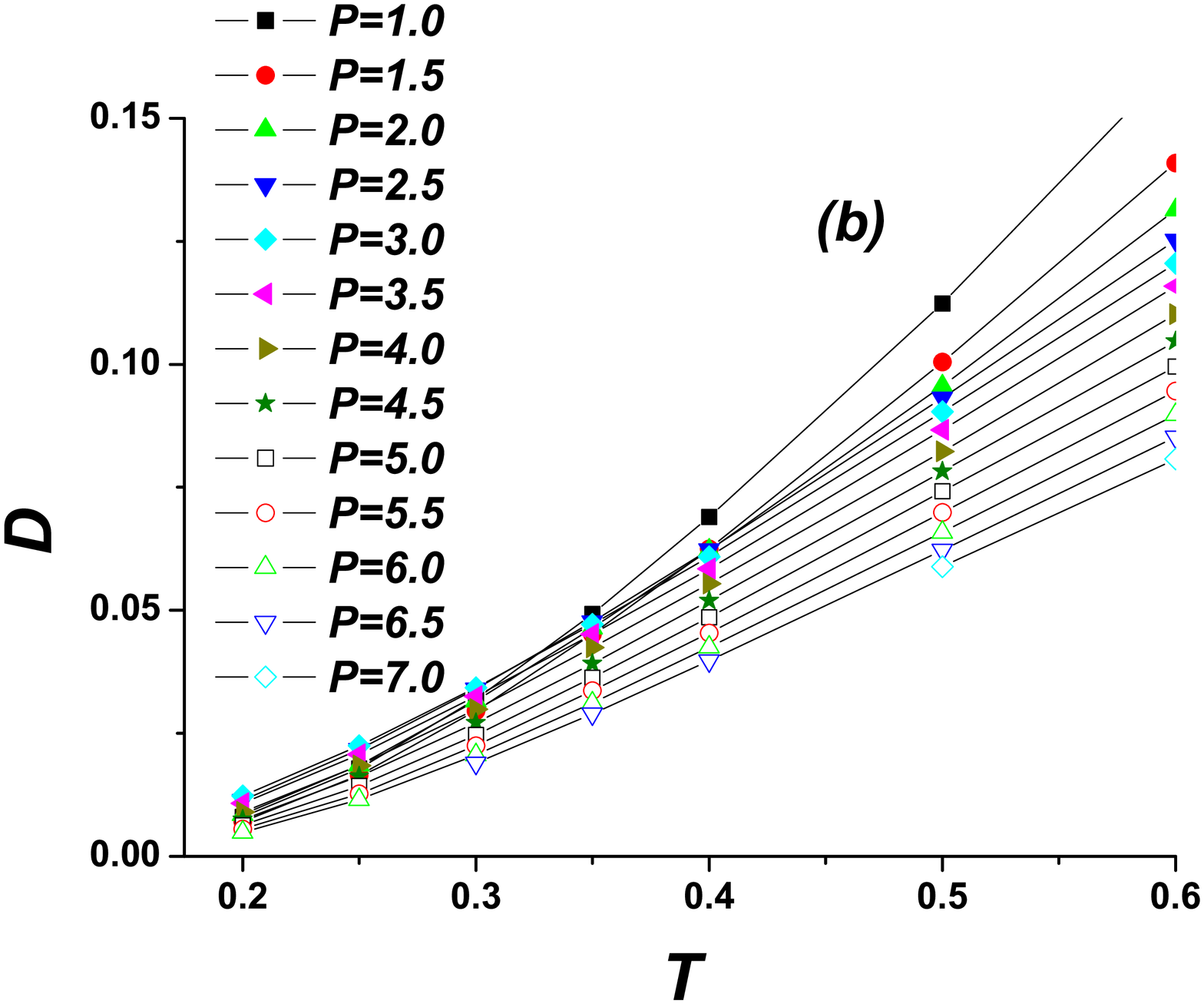}%
\caption{\label{fig:fig2n} (Color online). Diffusion coefficient
of the SRSS-AW system (system 3 in Table 1) along isobars as a
function of (a) density and (b) temperature.}
\end{figure}

Figs.~\ref{fig:fig2n}((a) and (b)) show the diffusion coefficient
along isobars as functions of density and temperature if
attraction is added (system 3 in Table 1). As in the case of
isotherms and isochores, the attraction makes the anomalies more
pronounced. For example, without attraction we could find only
hints of the diffusion anomaly as a function of density
(Fig.~\ref{fig:fig2} (a)), however, this anomaly is explicit for
the system 3 (see Fig.~\ref{fig:fig2n} (a)).

The last physically meaningful trajectory considered in the
present work is the adiabat. This trajectory is defined as
constant entropy curve. The entropy is calculated as following. We
compute excess free energy by integrating the equation of states:
$\frac{F_{ex}}{Nk_BT}=\frac{F-F_{id}}{Nk_BT}=\frac{1}{k_BT}
\int_0^{\rho} \frac{P(\rho ')-\rho ' k_BT}{\rho '^2} d\rho'$. The
excess entropy can be computed via $S_{ex}=\frac{U-F_{ex}}{N
k_BT}$. The total entropy is $S=S_{ex}+S_{id}$, where the ideal
gas entropy is
$\frac{S_{id}}{Nk_B}=\frac{3}{2}\ln(T)-\ln(\rho)+\ln(\frac{(2 \pi
mk_B)^{3/2}e^{5/2}}{h^3})$. The last term in this expression is
constant and is not accounted in our calculations.

\begin{figure}
\includegraphics[width=7cm, height=7cm]{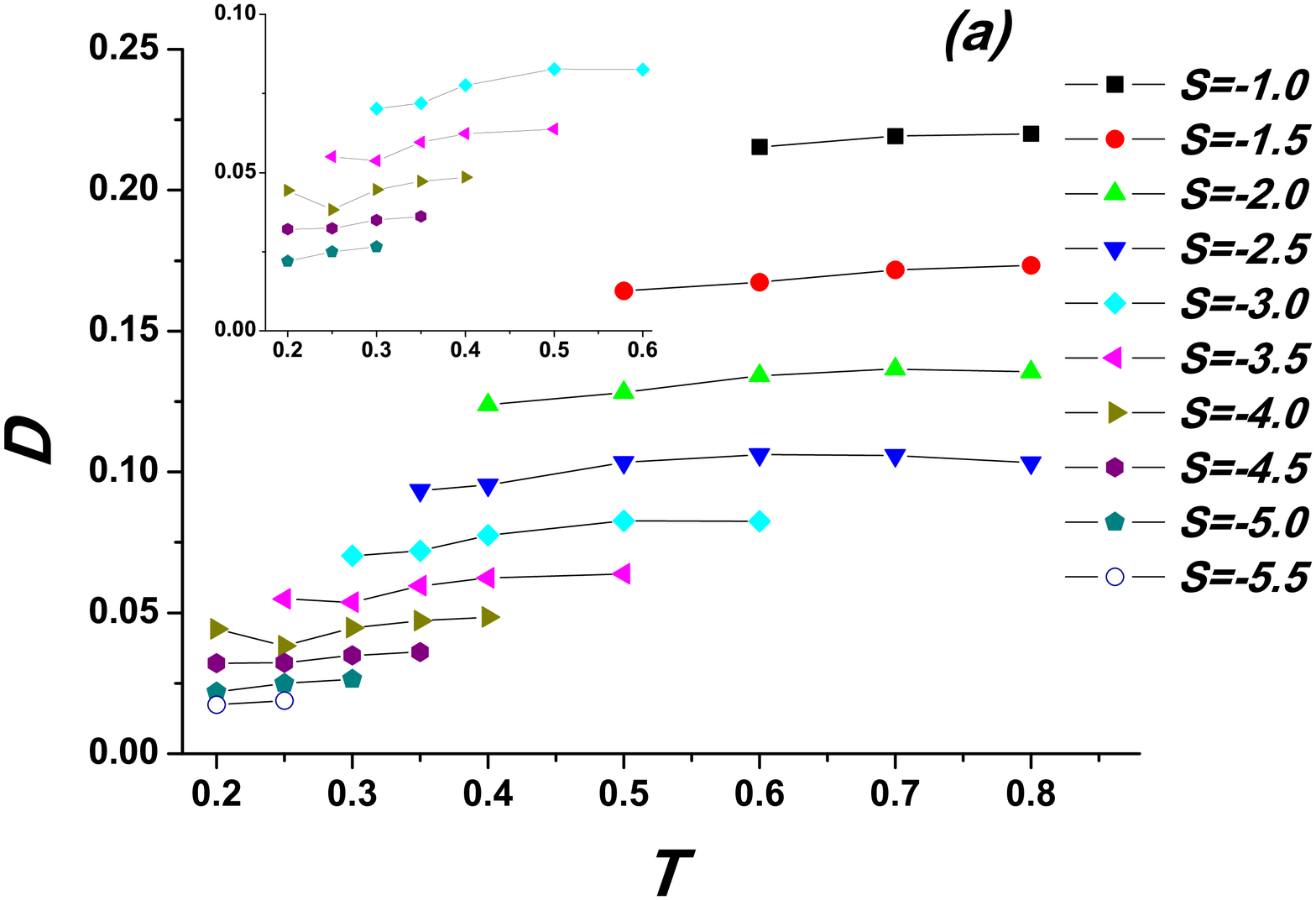}%

\includegraphics[width=7cm, height=7cm]{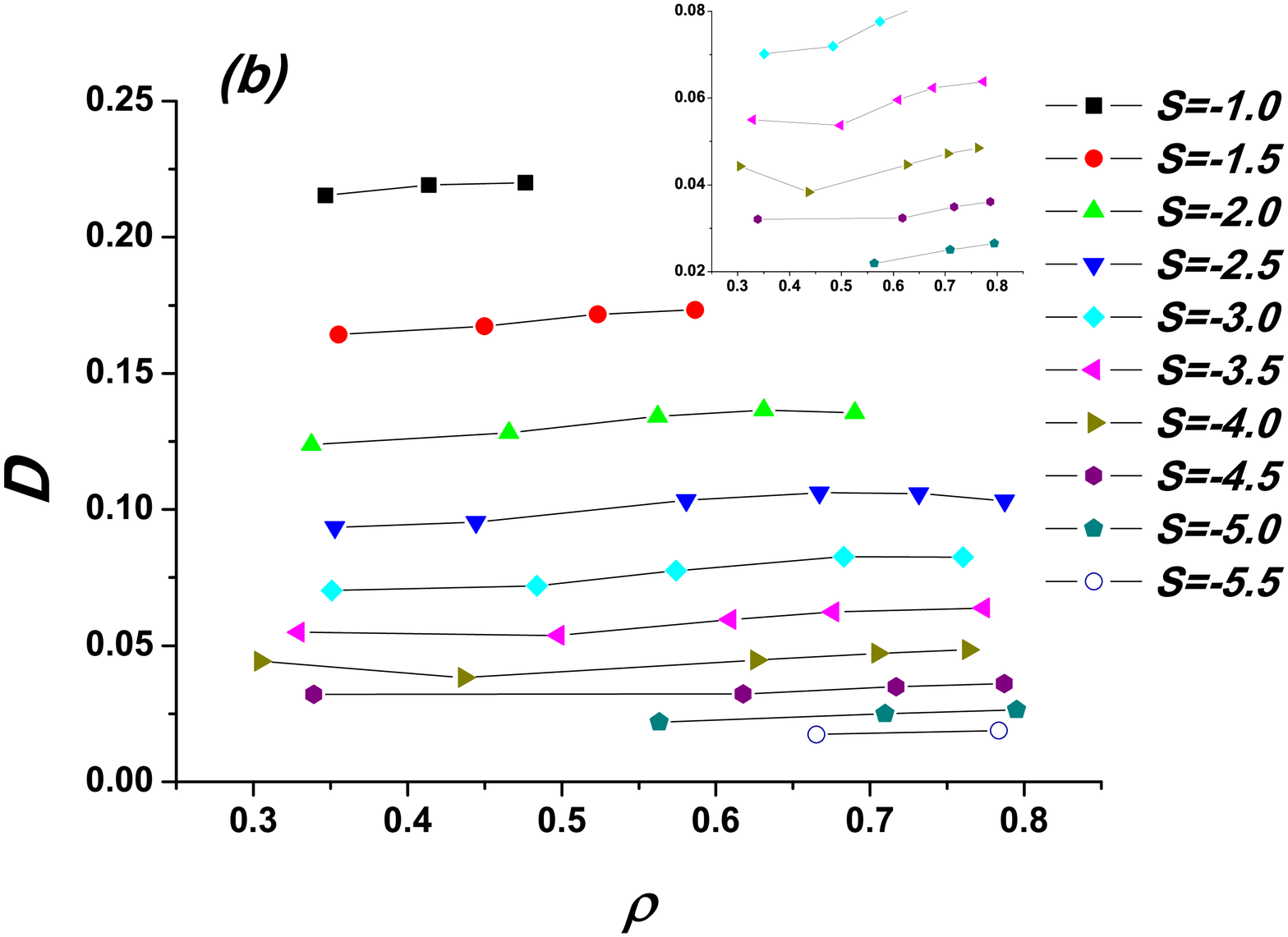}%

\includegraphics[width=7cm, height=7cm]{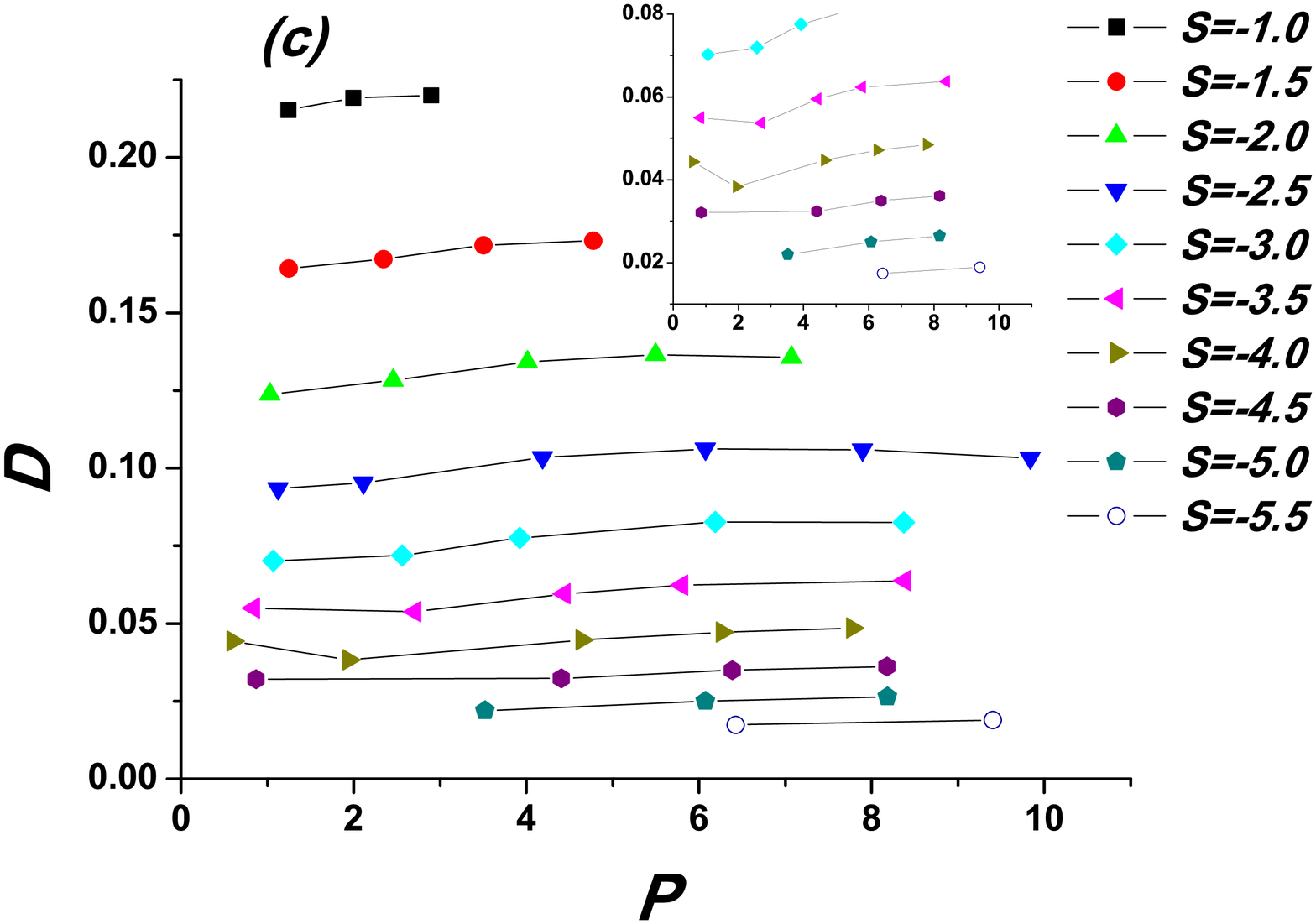}%
\caption{\label{fig:fig3} (Color online). Diffusion coefficient of
the SRSS system along adiabats as a function of (a) temperature,
(b) density, and (c) pressure. The inserts show anomalous regions
in the corresponding coordinates.}
\end{figure}

The behavior of entropy itself will be discussed below. Here we
give the diffusion coefficients along the adiabats
(Fig.~\ref{fig:fig3} (a)-(c)).

One can see from the Fig.~\ref{fig:fig3} (a)-(c) that the anomaly
takes place along adiabats in all possible coordinates (look, for
example, $S=-4.0$ adiabat). It means that in case of adiabatic
trajectory one can identify the anomalous region monitoring any of
three thermodynamic variables ($P,\rho,T$). However, this
trajectory is rather difficult to realize in simulation or
experiment.

As one can expect, the attraction does not produce qualitative
changes in the behavior of the anomalies, but makes them more
explicit. We do not show the corresponding figures.

\subsection{Density Anomaly}

\begin{figure}
\includegraphics[width=7cm, height=7cm]{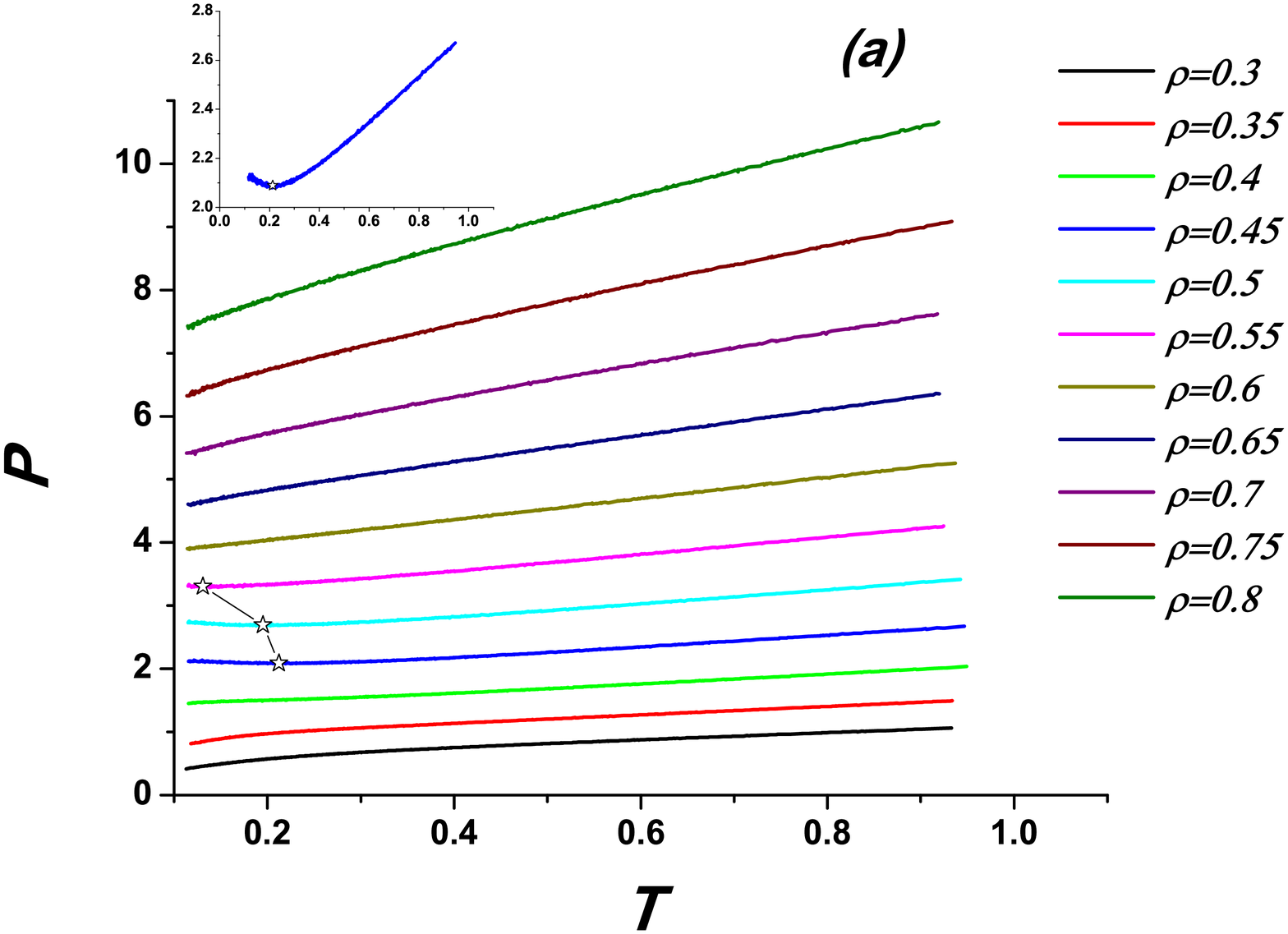}%

\includegraphics[width=7cm, height=7cm]{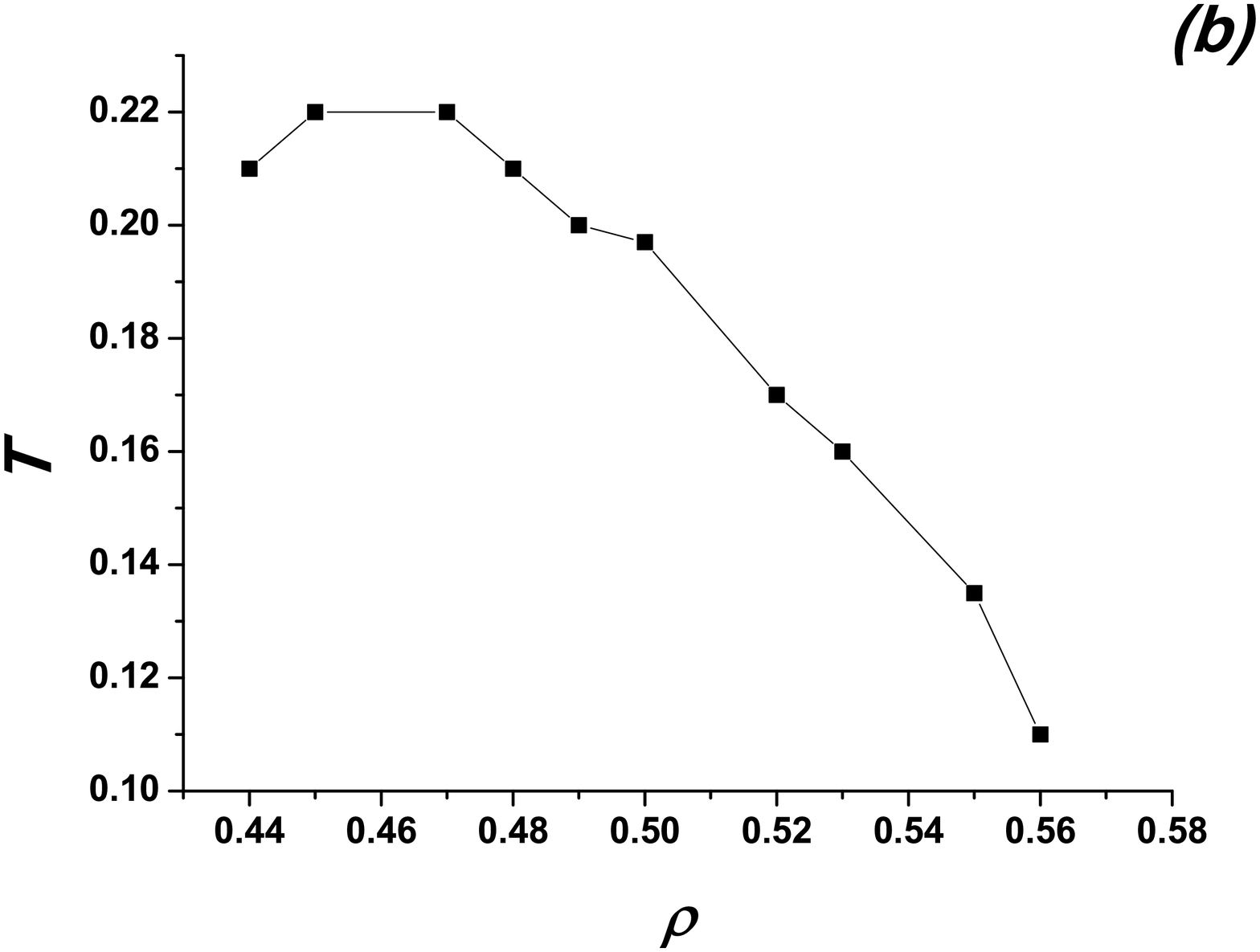}%

\caption{\label{fig:fig4} (Color online). (a) A set of isochors of
SRSS. The stars show the location of minimum. The insert enlarges
the $\rho=0.45$ isochor. (b) The location of the minima on
isochors in $\rho - T$ plane.}
\end{figure}

As we mentioned above density anomaly corresponds to appearance of
a minima on isochors of the system. The isochors of the system
with the potential (\ref{2}) (system 1 in Table 1) are shown in
the Fig.~\ref{fig:fig4}(a). It is evident from the figure that
some of the isochors do demonstrate minima. The location of the
minimum in the $\rho-T$ plane is shown in Fig.~\ref{fig:fig4}(b).

\begin{figure}
\includegraphics[width=7cm, height=7cm]{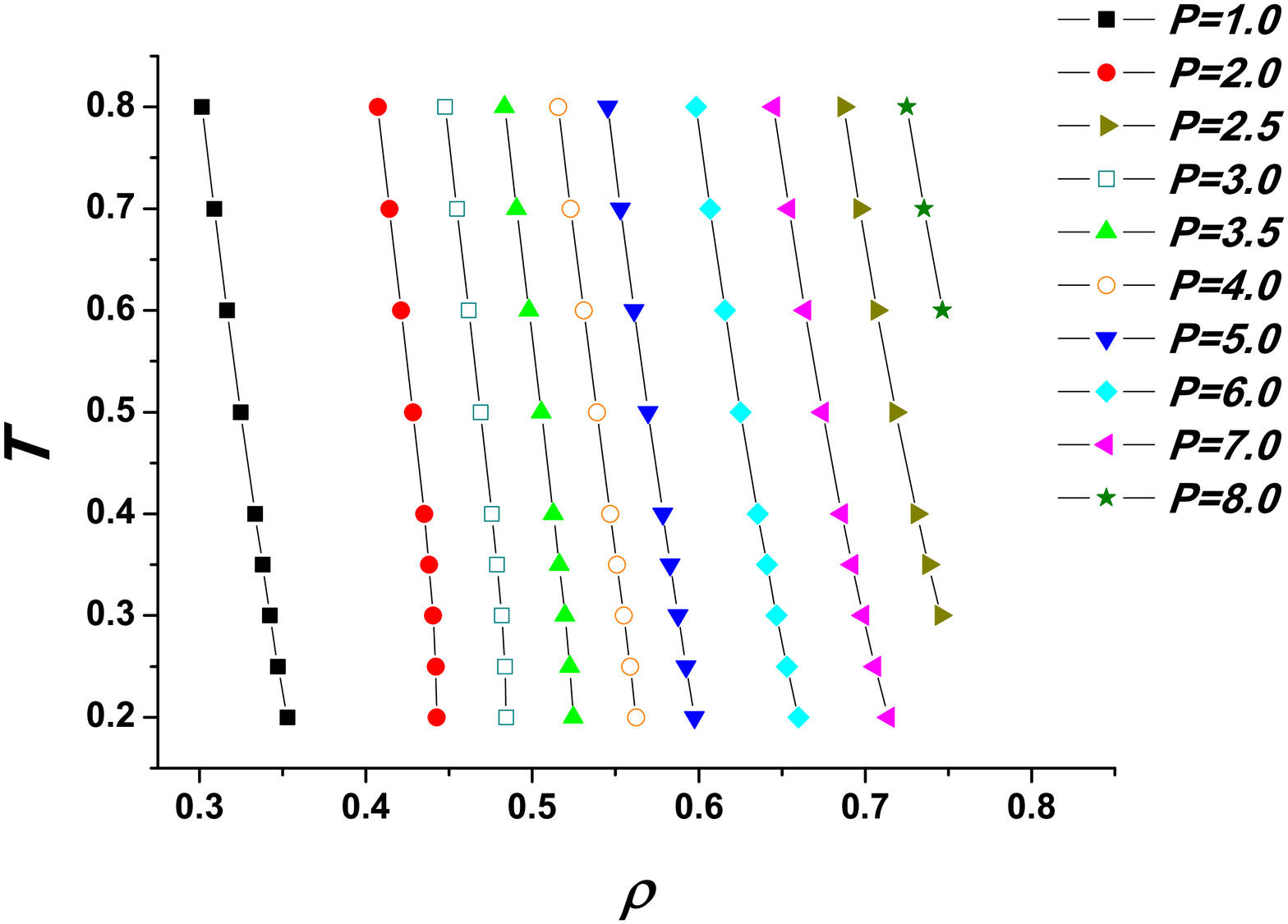}%

\caption{\label{fig:fig5} (Color online). A set of isobars of
SRSS. }
\end{figure}

If we turn to the shape of the isobars themselves for the
potential (\ref{2}) (system 1 in Table 1), that is the dependence
of temperature on density at fixed pressure, we do not find any
traces of anomalies there (Fig. ~\ref{fig:fig5}). Like in the case
of diffusion, the curves have negative slope which approaches zero
at low temperatures and densities corresponding to anomalous
regime. However, the curve remains monotonous, and it seems that
there is no sign of density anomaly along isobars. However, using
the well known thermodynamic relation:
\begin{equation}
\left(\frac{\partial V}{\partial T}\right)_P \left(\frac{\partial
T}{\partial P}\right)_V \left(\frac{\partial P}{\partial
V}\right)_T=-1, \label{td}
\end{equation}
one can find that $K_T\left(\frac{\partial P}{\partial T}\right)_V
\left(\frac{\partial T}{\partial
\rho}\right)_P=-\frac{N}{\rho^2}$, where $K_T$ is the isothermal
compressibility. Taking into account that $K_T$ is always positive
and finite for systems in equilibrium, and using
Fig.~\ref{fig:fig4}(a), one can see that $\left(\frac{\partial
T}{\partial \rho}\right)_P>0$ if $\left(\frac{\partial P}{\partial
T}\right)_V<0$. We can conclude that the anomaly does exist along
isobars for low temperatures, but we do not see it in our
simulations.

\begin{figure}
\includegraphics[width=7cm, height=7cm]{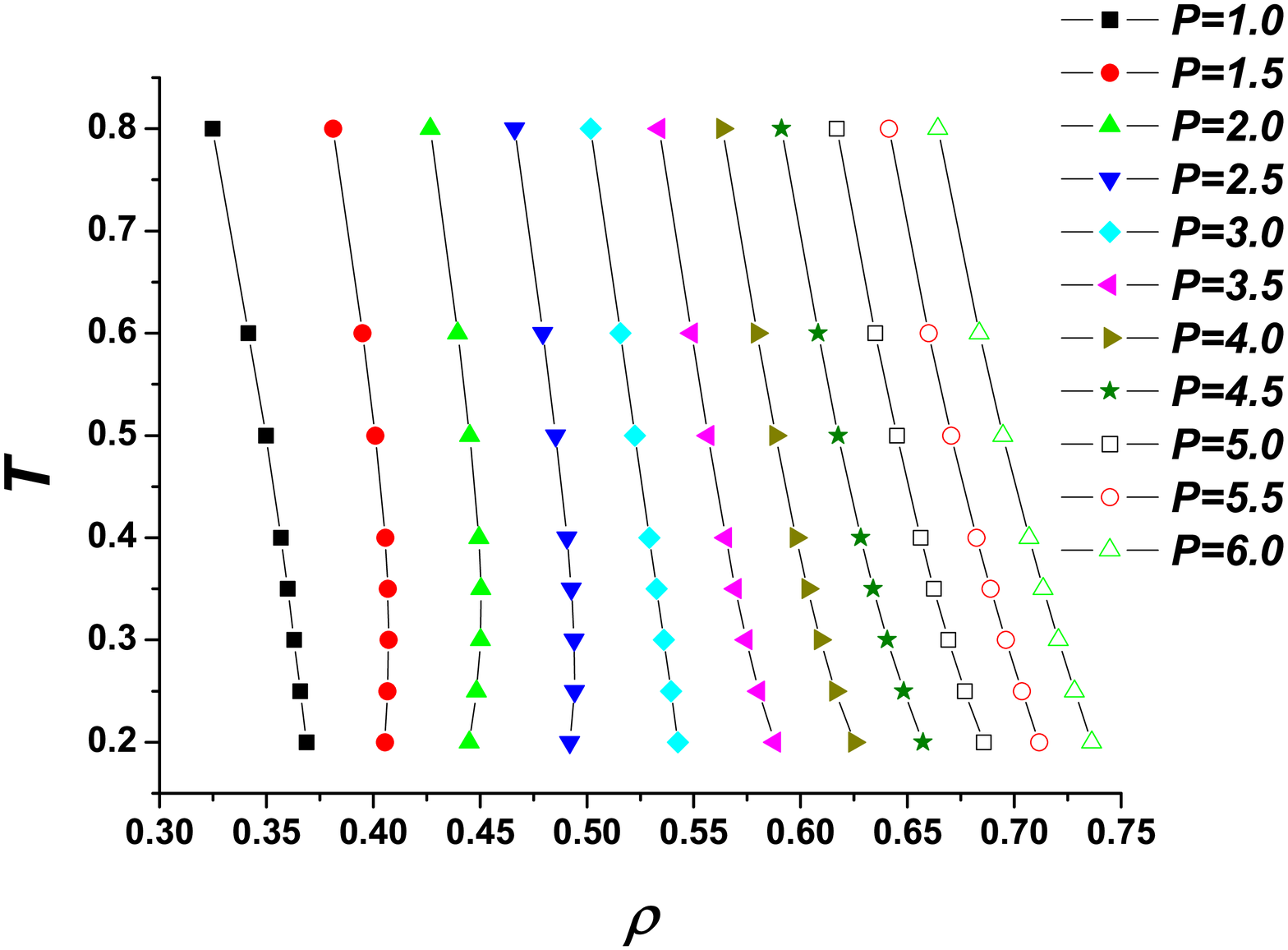}%

\caption{\label{fig:fig12} (Color online). A set of isobars of
SRSS-AW (system 3 in Table 1). }
\end{figure}

As it was mentioned above, adding the attraction to the potential
makes the anomalies more pronounced. To illustrate this, in
Fig.~\ref{fig:fig12} we show the isobars for the system 3 (see
Table 1) where the anomalies are clear seen.

\begin{figure}
\includegraphics[width=7cm, height=7cm]{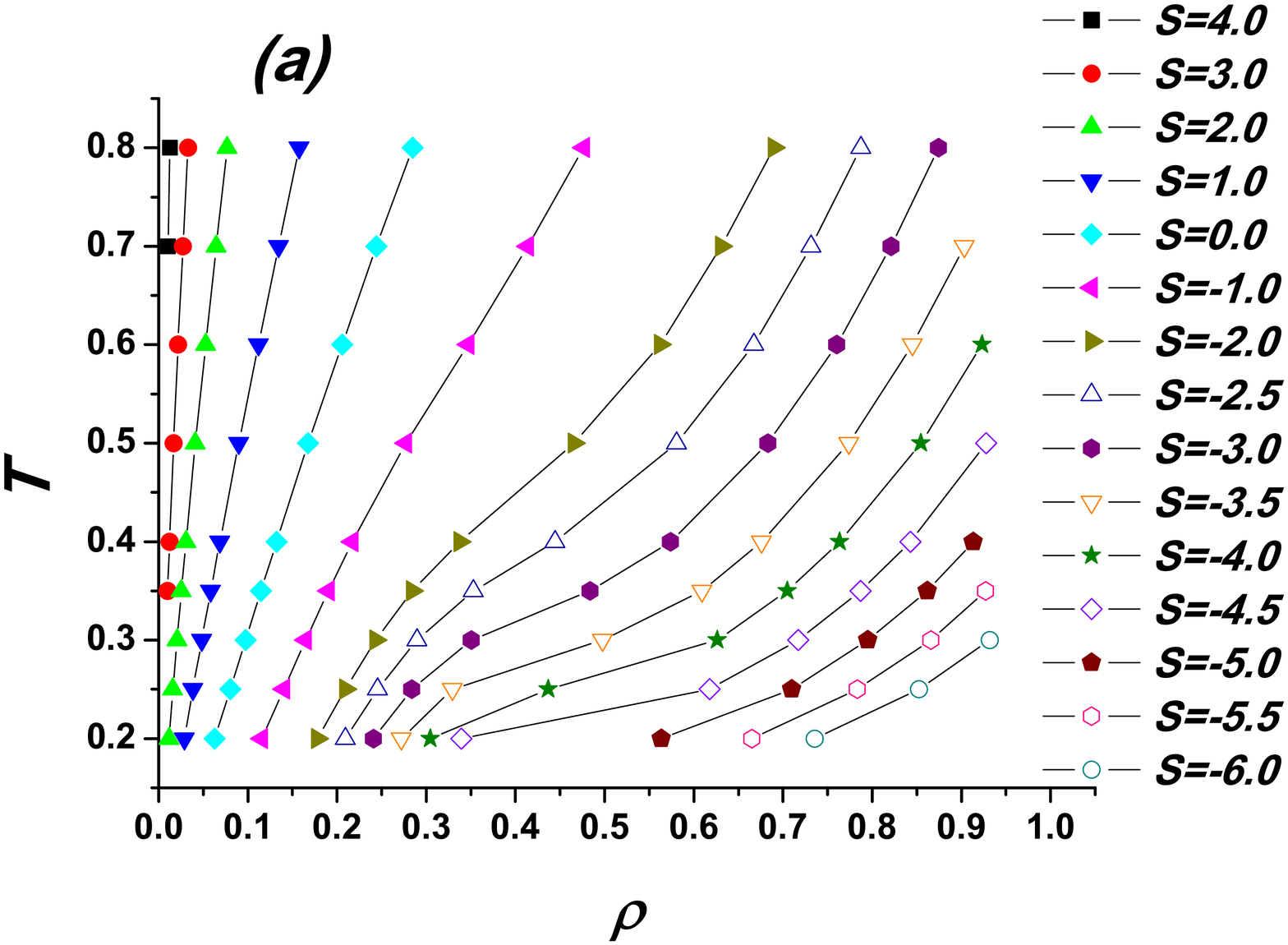}%

\includegraphics[width=7cm, height=7cm]{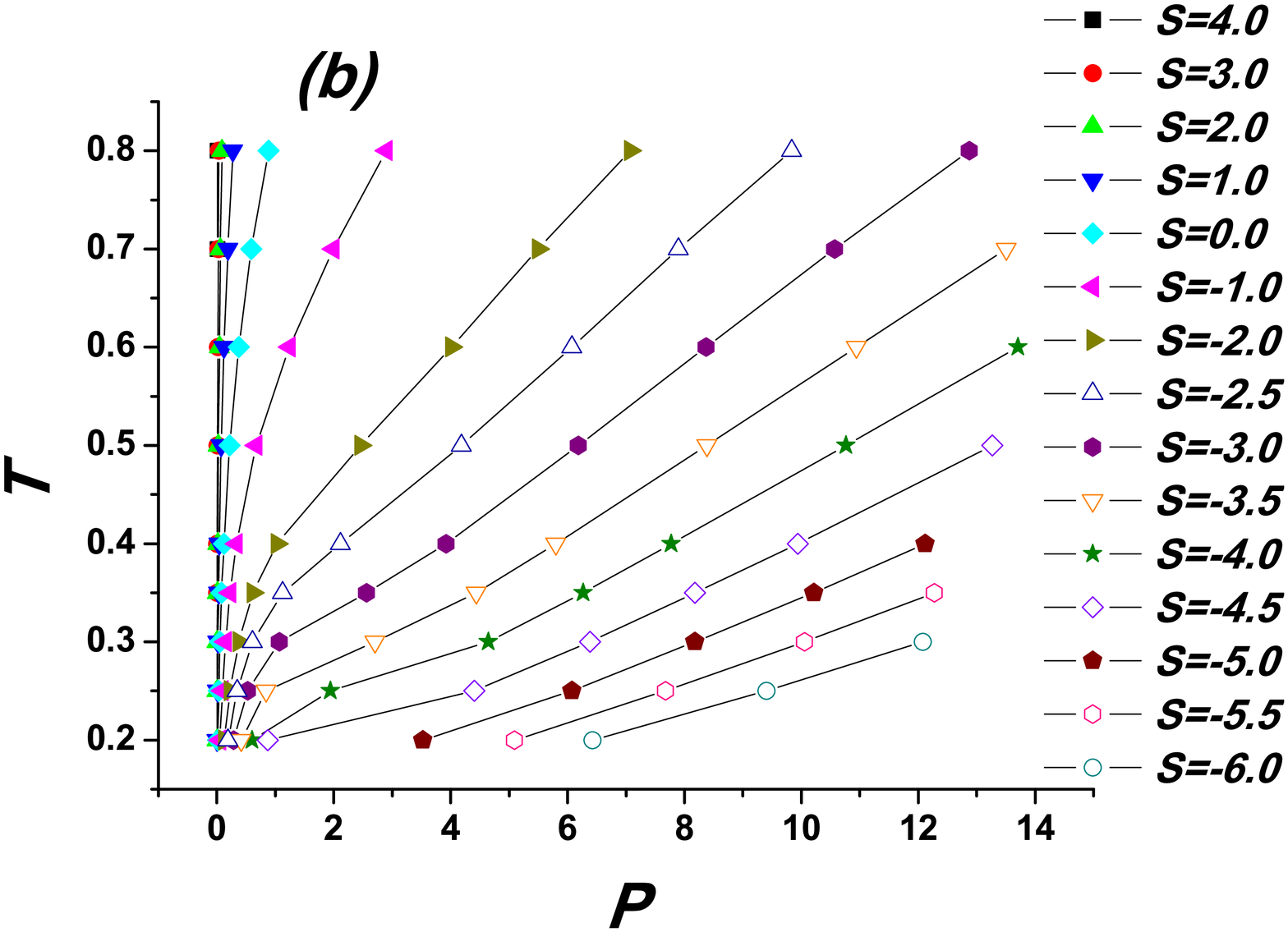}%

\caption{\label{fig:fig6} (Color online). Adiabats of the SRSS in
(a) $\rho-T$ and (b) $P-T$ coordinates.}
\end{figure}

Figs.~\ref{fig:fig6} ((a) and (b)) show adiabats of the SRSS in
$\rho-T$ and $P-T$ coordinates. Interestingly, the curves seem
monotonic, but the slope of the curves at low and high densities
(pressures) is very different. At the same time the middle density
(pressure) curves demonstrate a continues change from the high
slope (low-density regime) to low slope (high-density regime).
However, from the well-known thermodynamic relation
$\left(\frac{\partial T}{\partial
\rho}\right)_S=\rho^2\frac{T}{c_v}\left(\frac{\partial P}{\partial
T}\right)_V$ and Fig.~\ref{fig:fig4} one can see that in the case
of Fig.~\ref{fig:fig6} (a) the anomaly does exists, but it is not
seen because of the insufficient accuracy of calculations along
the adiabats. On the other hand, due to the relation
$\left(\frac{\partial T}{\partial
P}\right)_S=\frac{T}{c_P}\left(\frac{\partial V}{\partial
T}\right)_P$ there is also anomalous behavior in
Fig.~\ref{fig:fig6} (b).

As one can expect (we do not represent these figures here for the
sake of brevity), the anomalies are much better seen for the
systems with the attractive potentials (systems 2 and 3 in Table
1).

\subsection{Structural Anomaly}

Structural anomaly region can be bounded by using the local order
parameters or by excess entropy minimum and maximum. Here we apply
the definition via excess entropy.

The behavior of excess entropy is qualitatively analogous to the
behavior of diffusion coefficient. Because of this we briefly
describe it here noting that most of the conclusions about
diffusion coefficient along different trajectories can be applied
to excess entropy as well.

\begin{figure}
\includegraphics[width=7cm, height=7cm]{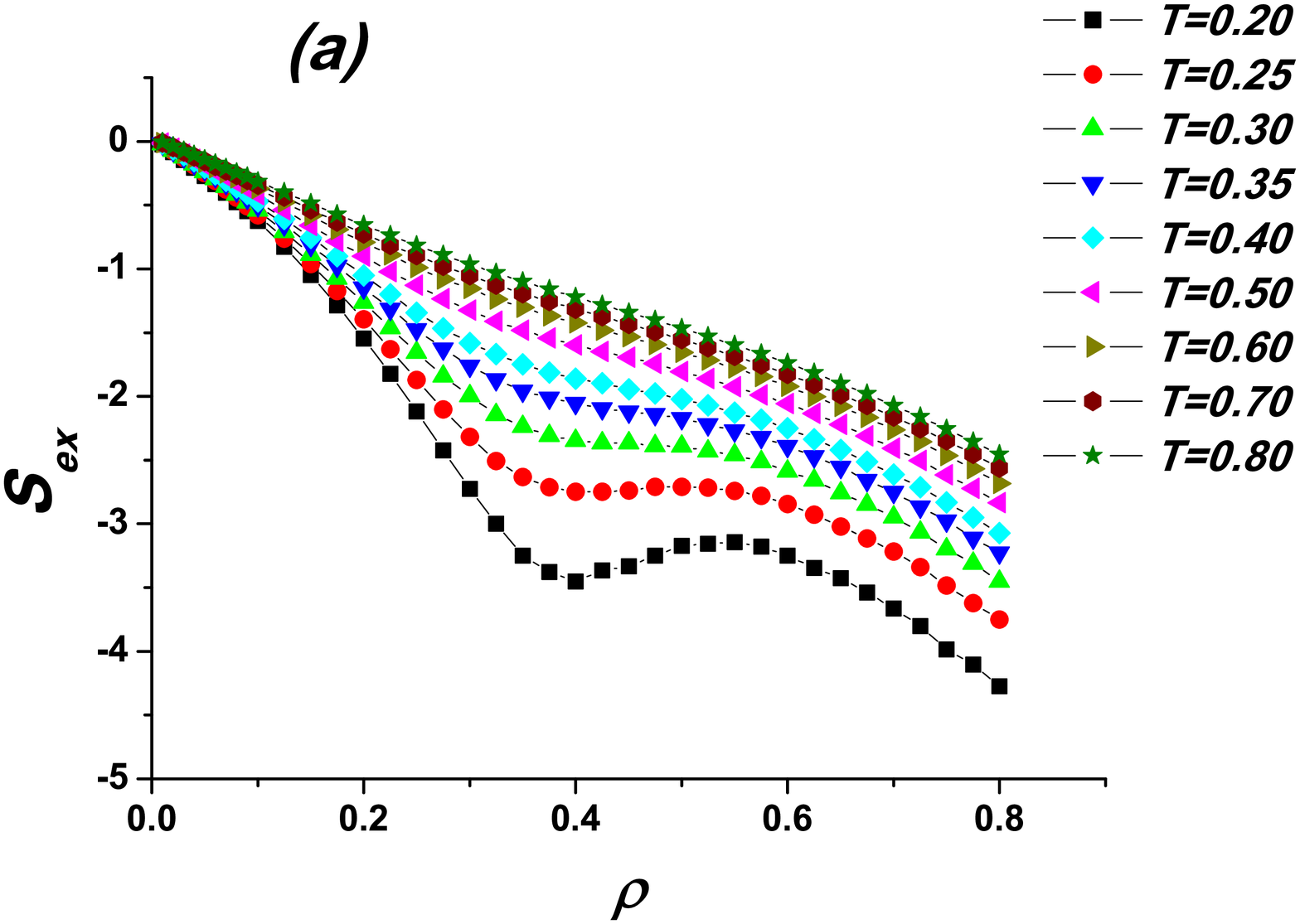}%

\includegraphics[width=5cm, height=7cm]{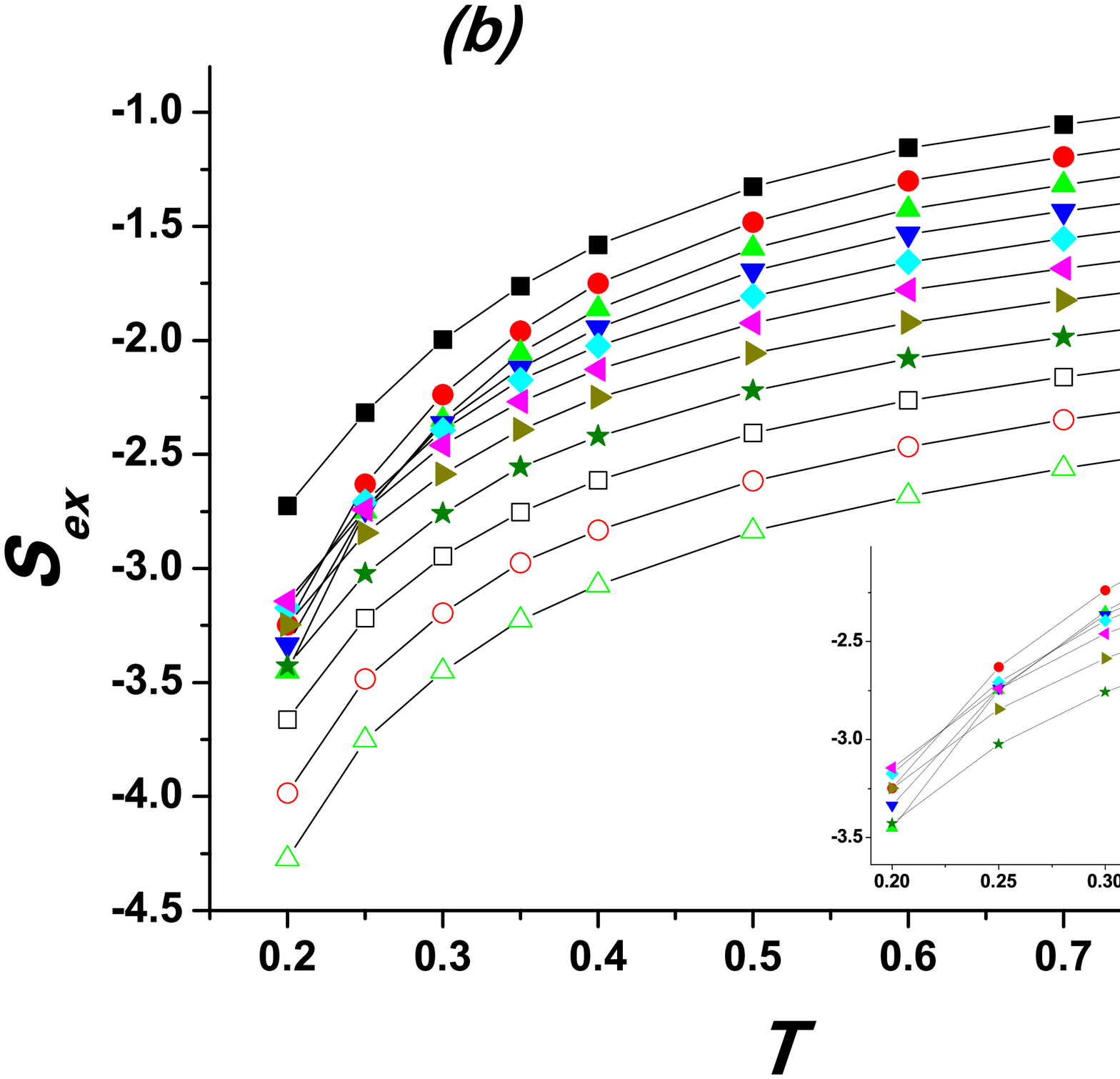}%

\caption{\label{fig:fig7} (Color online). Excess entropy of SRSS
fluid along (a) isotherms and (b) isochors.}
\end{figure}

Figs.~\ref{fig:fig7} ((a) and (b)) show the excess entropy along
isotherms and isochors for the purely repulsive potential
(\ref{2}). Like for the diffusion coefficient, excess entropy
demonstrates anomalous grows in some density range at low
temperatures. At the same time excess entropy is monotonous along
isochors. However, the curves for different isochors cross which
indicates the presence of anomaly.

\begin{figure}
\includegraphics[width=7cm, height=7cm]{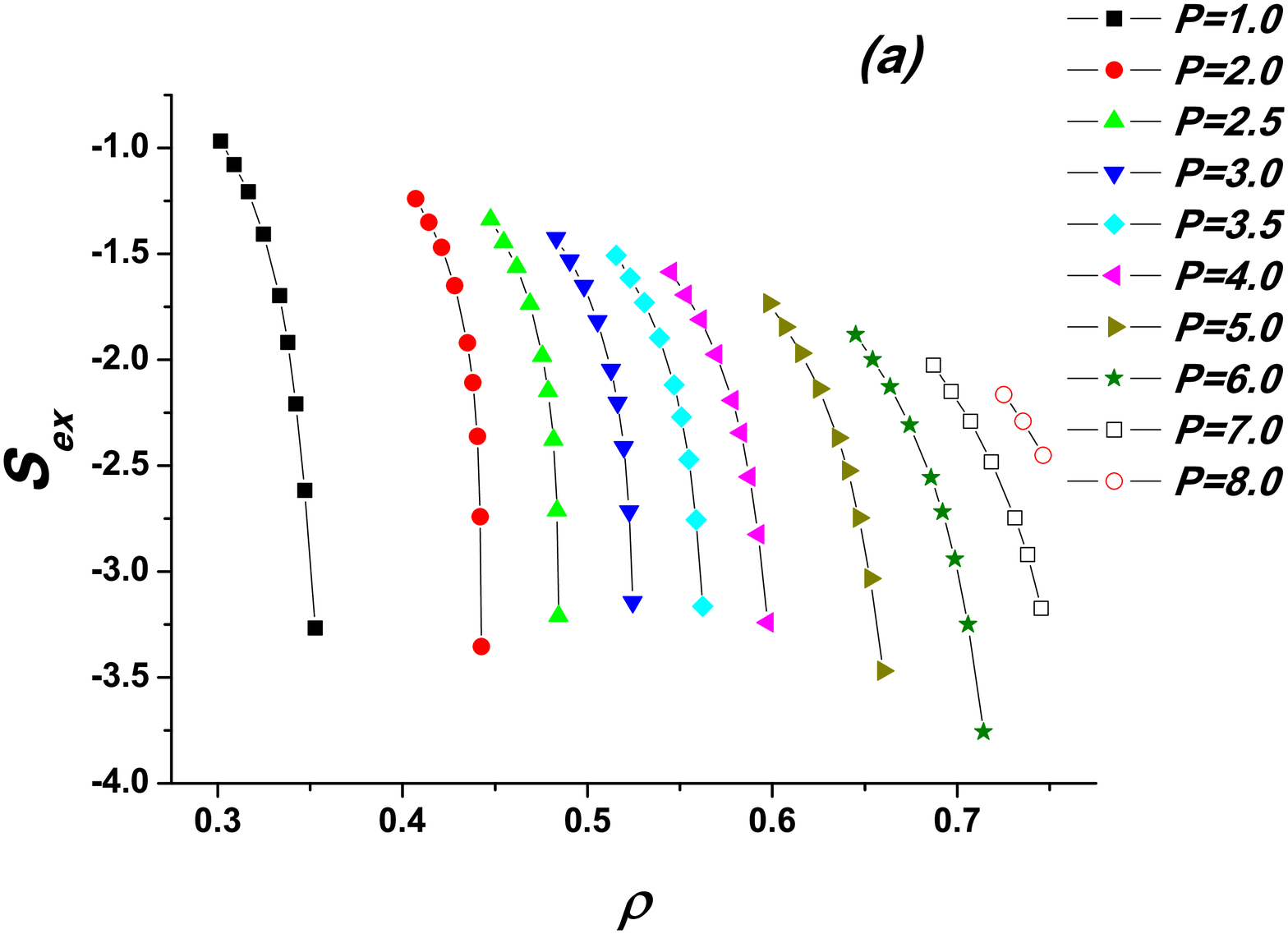}%

\includegraphics[width=7cm, height=7cm]{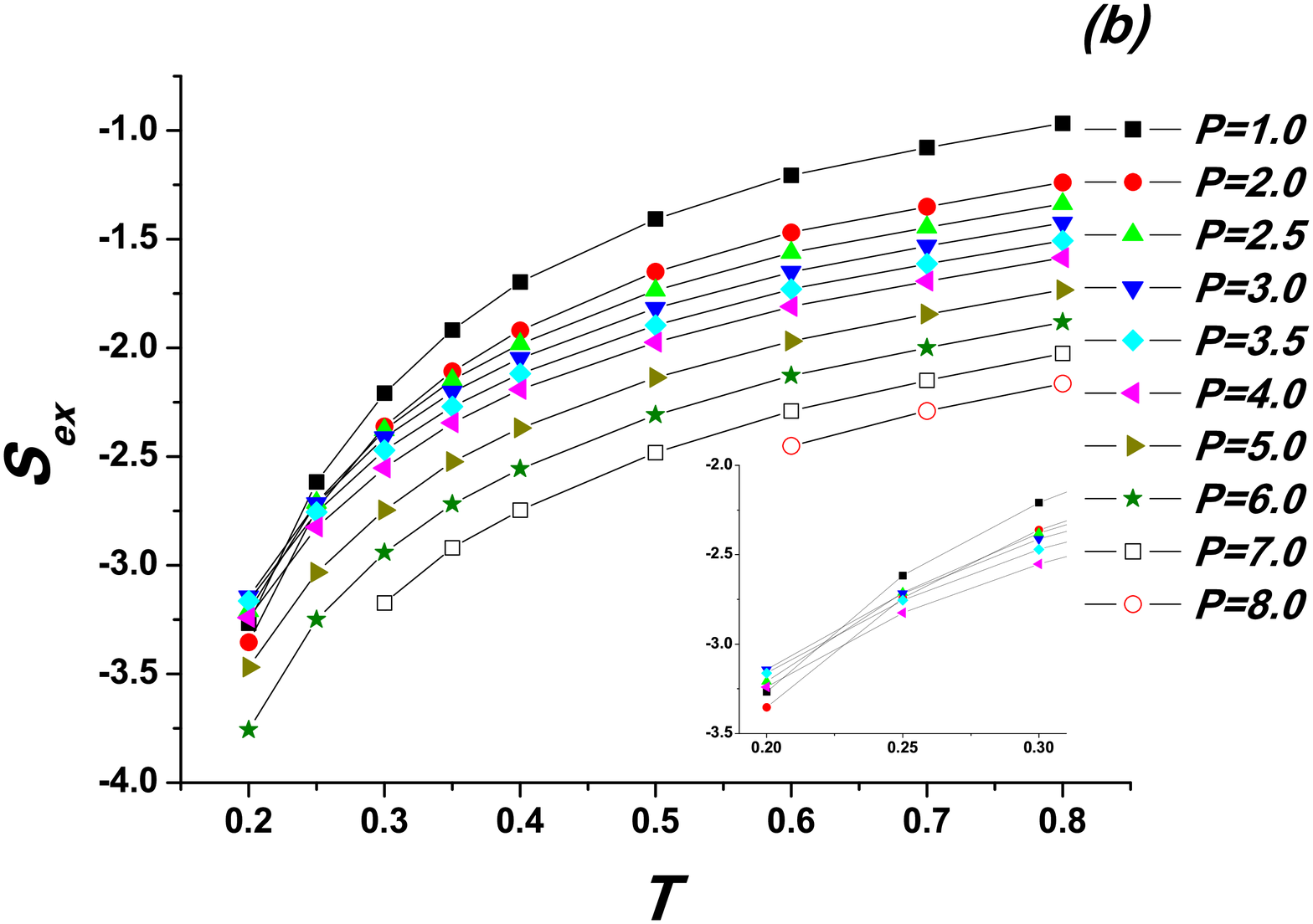}%

\caption{\label{fig:fig8} (Color online). Excess entropy of SRSS
fluid along isobars as a function of (a) density and (b)
temperature. The insert in (b) shows the cross of the curves at
low temperatures.}
\end{figure}

\begin{figure}
\includegraphics[width=7cm, height=7cm]{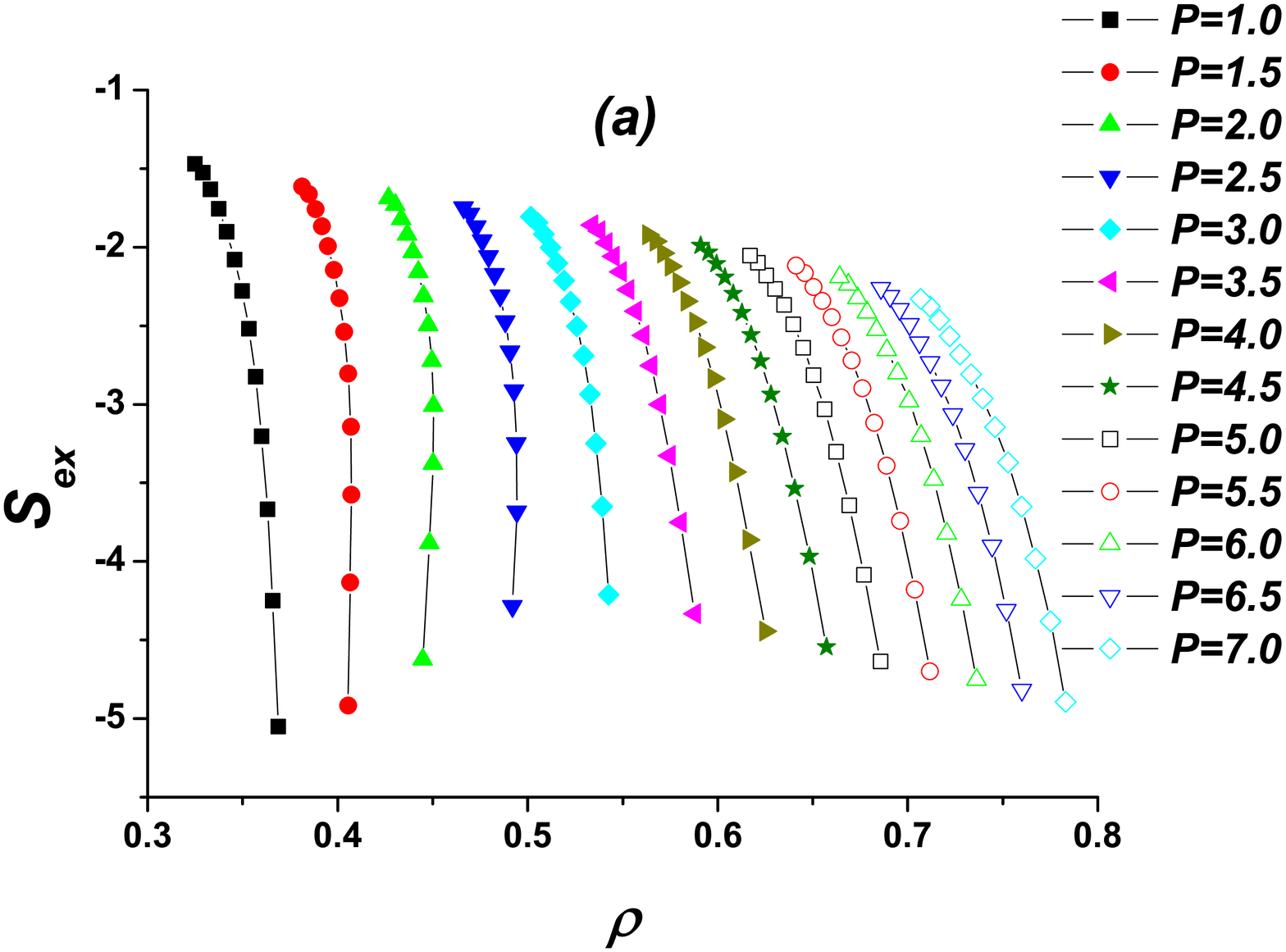}%

\includegraphics[width=7cm, height=7cm]{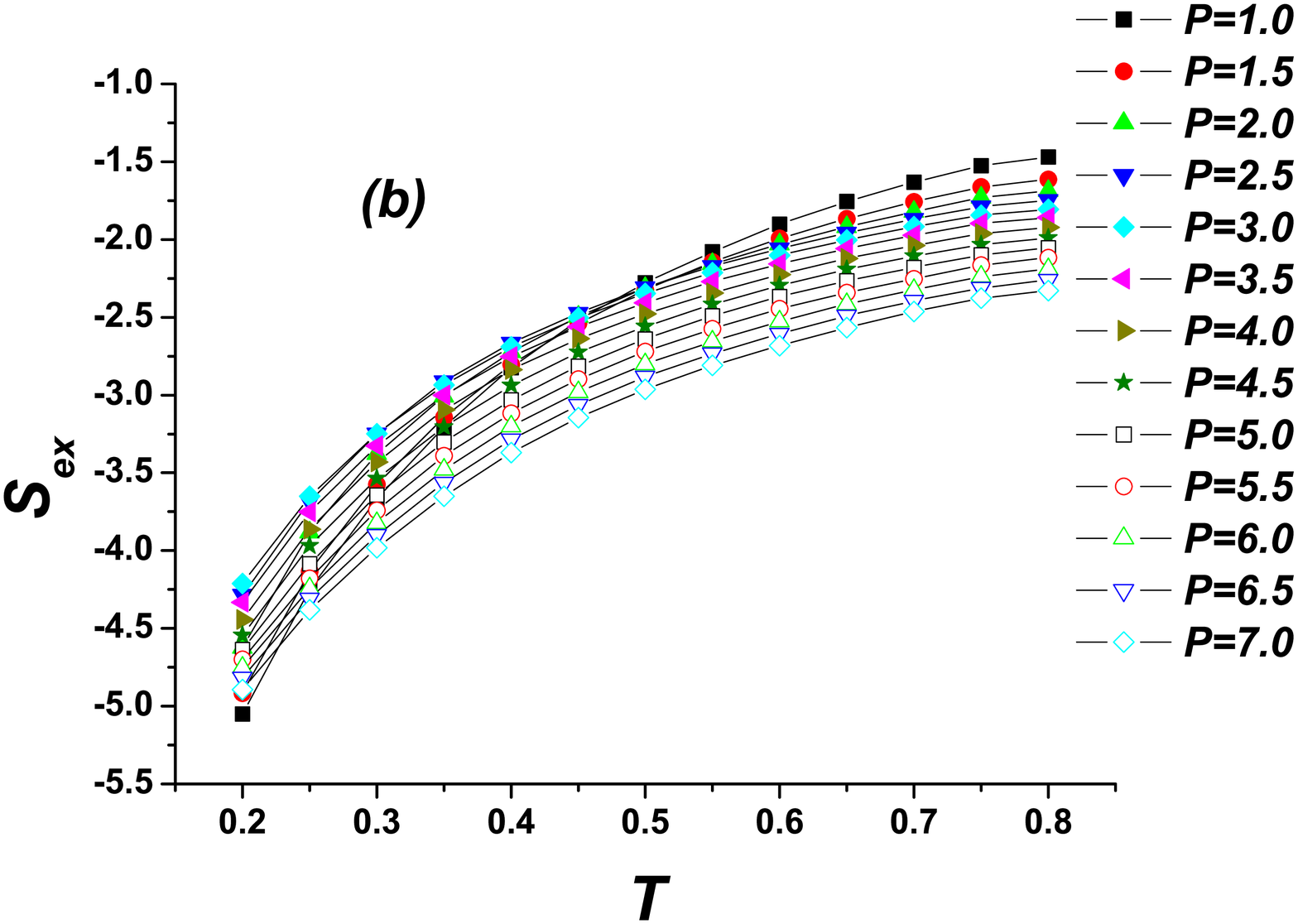}%

\caption{\label{fig:fig15} (Color online). Excess entropy of
SRSS-AW fluid along isobars as a function of (a) density and (b)
temperature (system 3 in Table 1).}
\end{figure}

The excess entropy along isobars is monotonically decreasing
function of density and monotonically increasing function of
temperature (see Figs.~\ref{fig:fig8}((a) and (b))). However, the
curves cross at low temperatures indicating the presence of
anomalies as it was discussed for the case of diffusion. As in the
case of diffusion anomaly, the structural anomaly is not seen
along the isobars for the purely repulsive potential (system 1 in
Table 1), however, one can expect that in our simulation we could
not reach the anomalous region. If the attraction is added to the
potential, the anomaly becomes explicit even as a function of
density (see Figs.~\ref{fig:fig15}((a) and (b))).

\begin{figure}
\includegraphics[width=7cm, height=7cm]{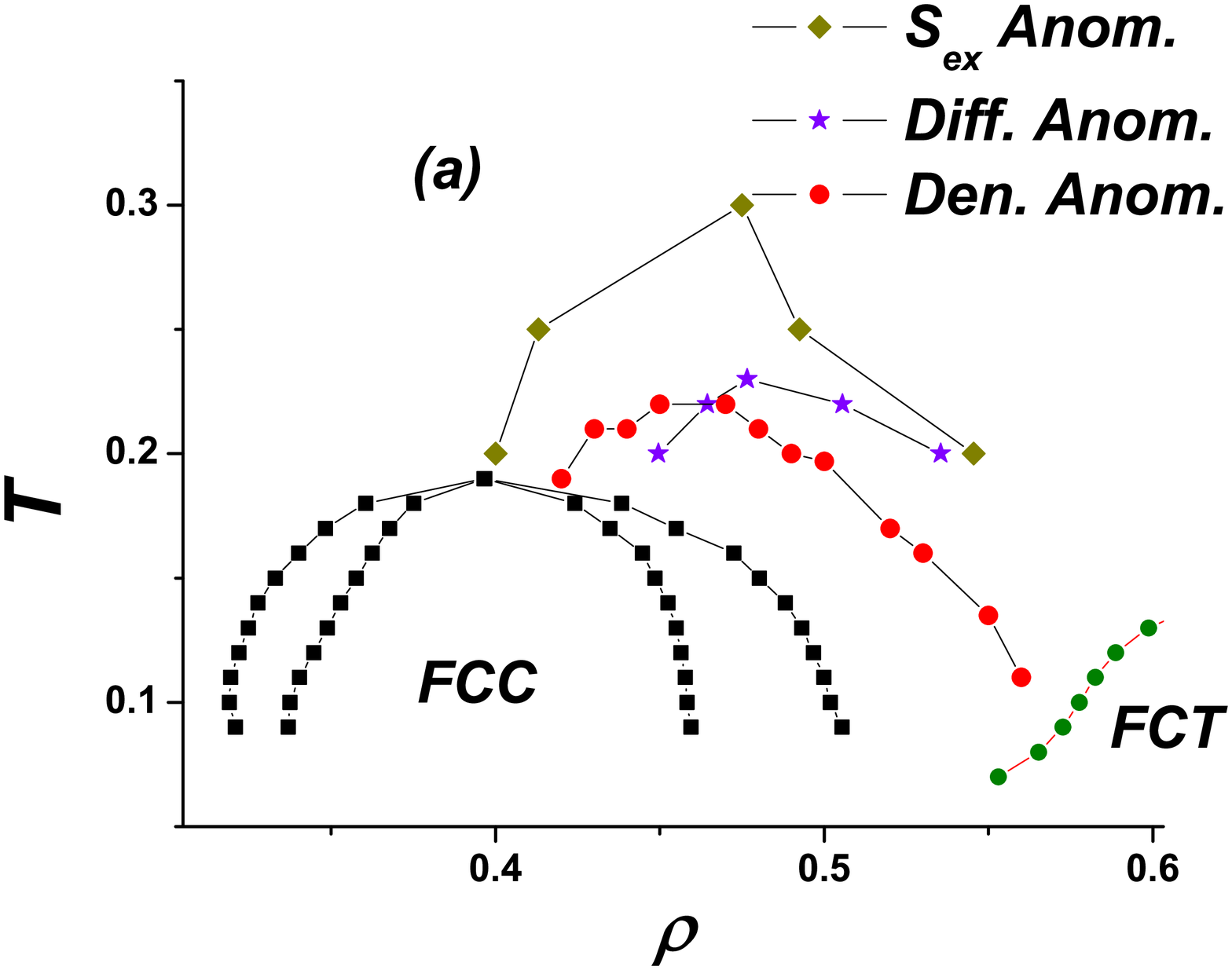}%

\includegraphics[width=7cm, height=7cm]{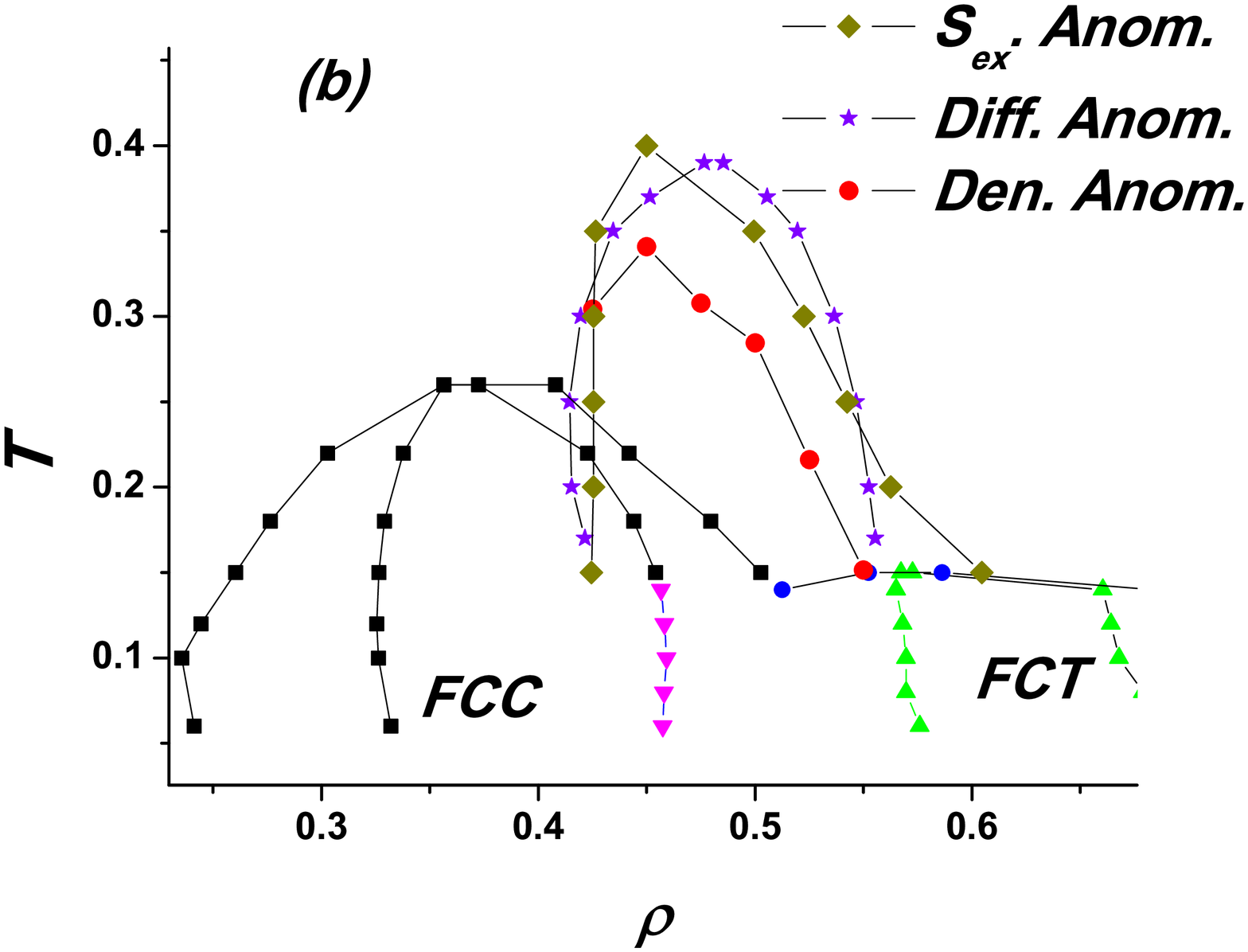}%

\includegraphics[width=7cm, height=7cm]{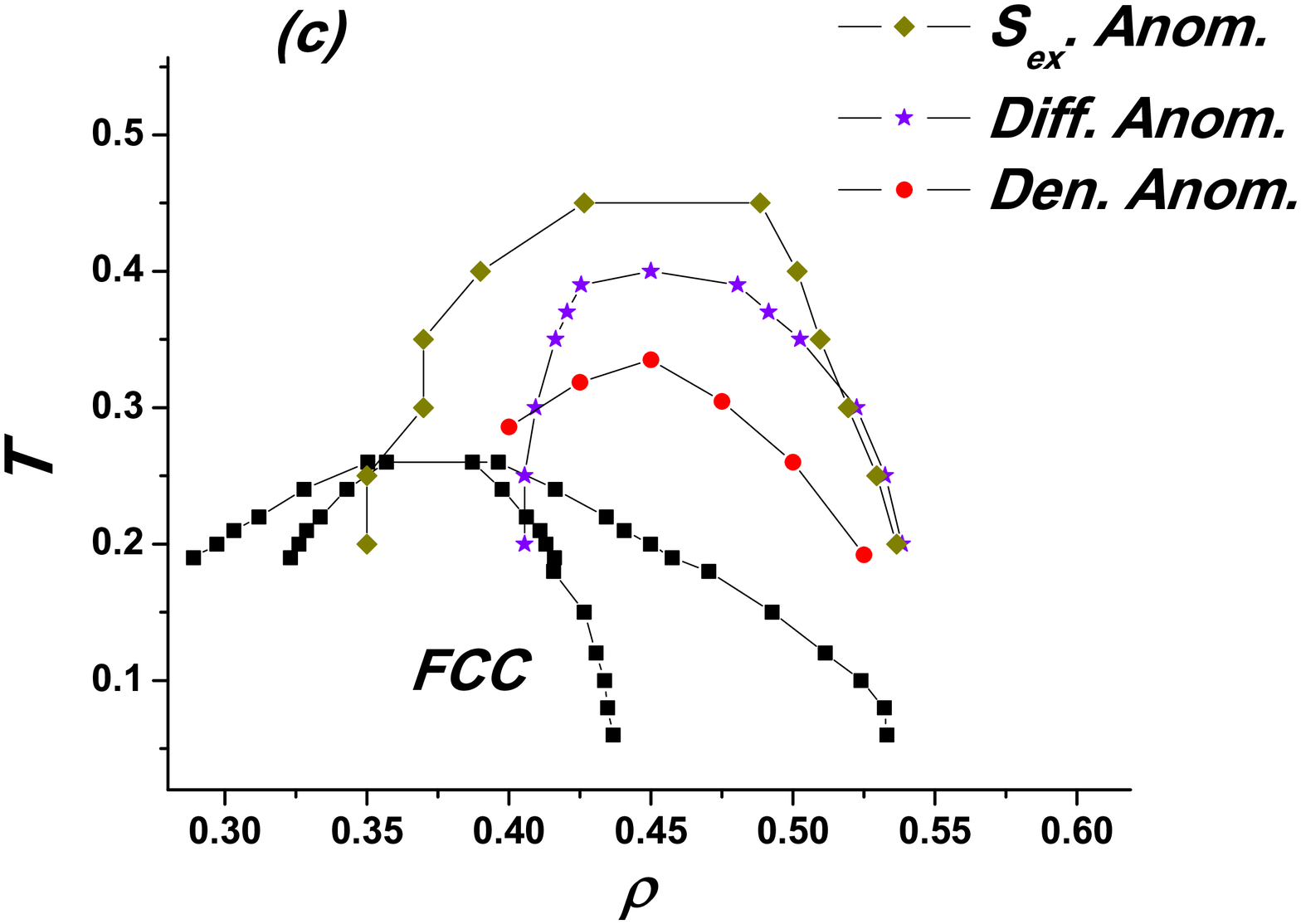}%

\caption{\label{fig:fig16} (Color online). Locations of the
anomalies lines on the phase diagrams in $T-\rho$ plane for the
system 1 (a), system (2) (b), and System 3 (c) from Table 1.}
\end{figure}

It is important to note that the range of densities of structural
anomalies is wider then the one of diffusion and density anomalies
which is consistent with the literature data for core-softened
systems \cite{deben2001,netz}. Figs.~\ref{fig:fig16}((a), (b) and
(c)) represent the locations of the anomalies lines on the phase
diagrams of the systems in Table 1 in $\rho-T$ plane.

\section{IV. Rosenfeld scaling}

In 1977 Rosenfeld proposed a connection between thermodynamic and
dynamical properties of liquids \cite{ros1,ros2}. The main
Rosenfeld's statement claims that the transport coefficients are
exponential functions of the excess entropy. In order to write the
exponential relations Rosenfeld introduced reduction of the
transport coefficients by some macroscopic parameters of the
system. For the case of diffusion coefficient one writes: $D^*=D
\frac{\rho ^{1/3}}{(k_BT/m)^{1/2}}$, where $m$ is the mass of the
particles. The Rosenfeld scaling rule can be written as:

\begin{equation}
 D^*=A \cdot e^{BS_{ex}},
\end{equation}
where $A$ and $B$ are constants.

In his original works Rosenfeld considered hard spheres, soft
spheres, Lennard-Jones system and one-component plasma
\cite{ros1,ros2}. After that the excess entropy scaling was
applied to many different systems including core-softened liquids
\cite{errington,errington2,india1,indiabarb,weros}, liquid metals
\cite{liqmet1,liqmet2}, binary mixtures \cite{binary1,binary2},
ionic liquids \cite{india2,ionicmelts}, network-forming liquids
\cite{india1,india2}, water \cite{buldwater}, chain fluids
\cite{chainfluids} and bounded potentials
\cite{weros,klekelberg,klekelberg1}.

In our recent publication \cite{weros,FR2011,werostr} we showed
that for the case of the core-softened fluids the applicability of
Rosenfeld relation depends on the trajectory. In particular,
Rosenfeld relation is applicable along isochors, but it is not
applicable along isotherms.

The breakdown of the Rosenfeld relation along isotherms can be
seen from the following speculation. The regions of different
anomalies do not coincide with each other. In particular, in the
case of core-softened fluids the diffusion anomaly region is
located inside the structural anomaly one. It means that there are
some regions where the diffusion is still normal while the excess
entropy is already anomalous. But this kind of behavior can not be
consistent with the Rosenfeld scaling law.

\begin{figure}
\includegraphics[width=7cm, height=7cm]{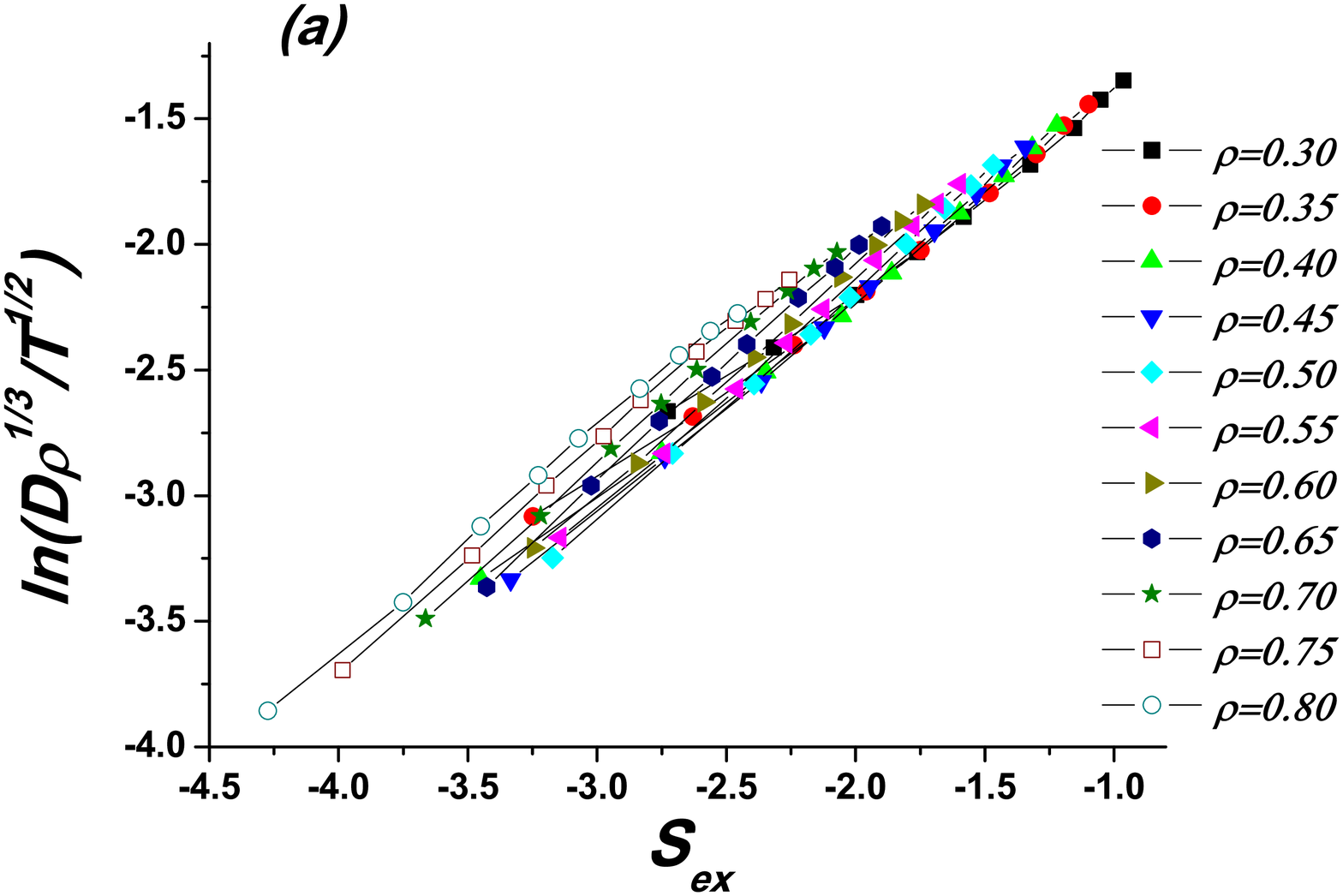}%

\includegraphics[width=7cm, height=7cm]{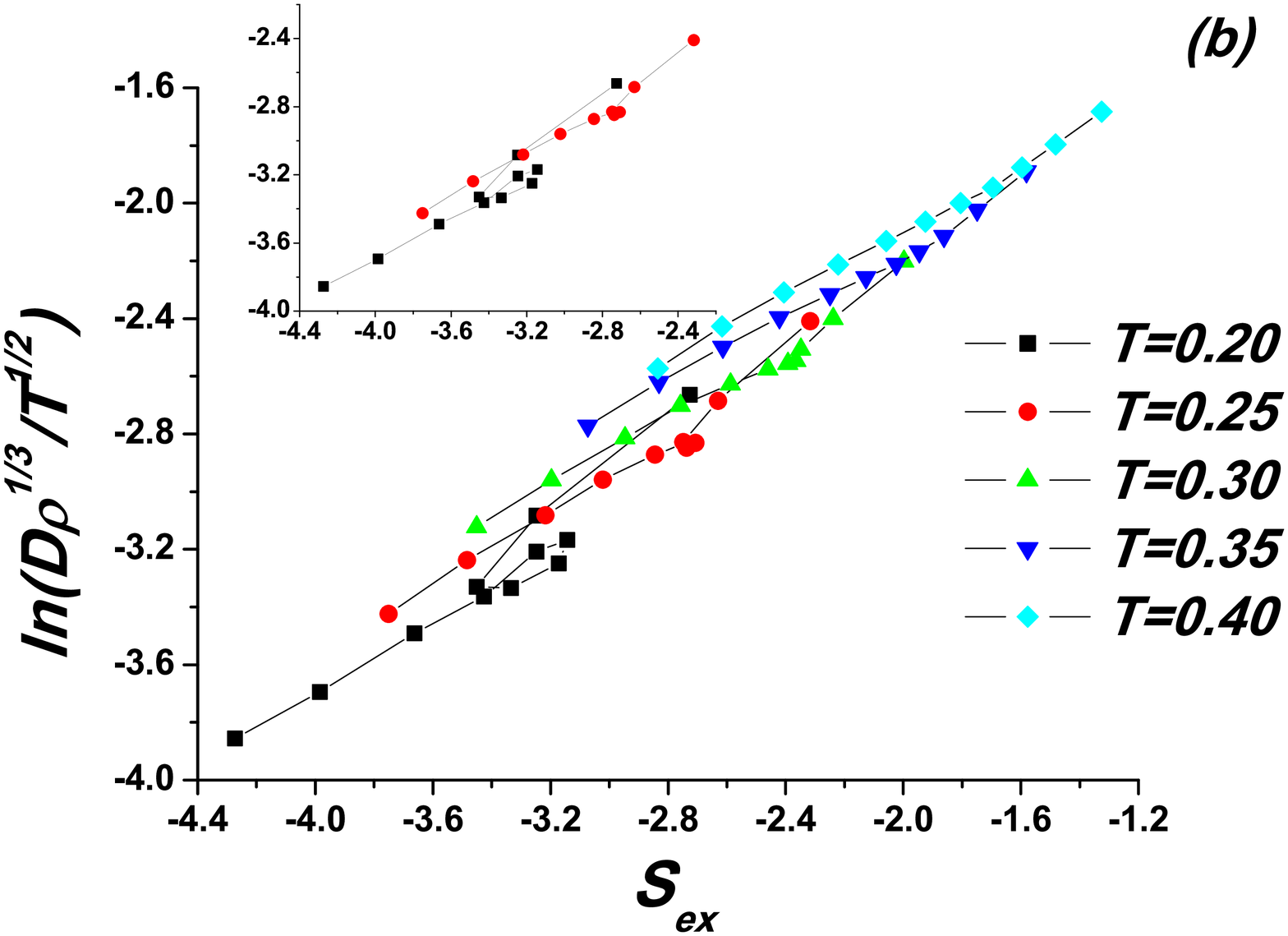}%

\includegraphics[width=7cm, height=7cm]{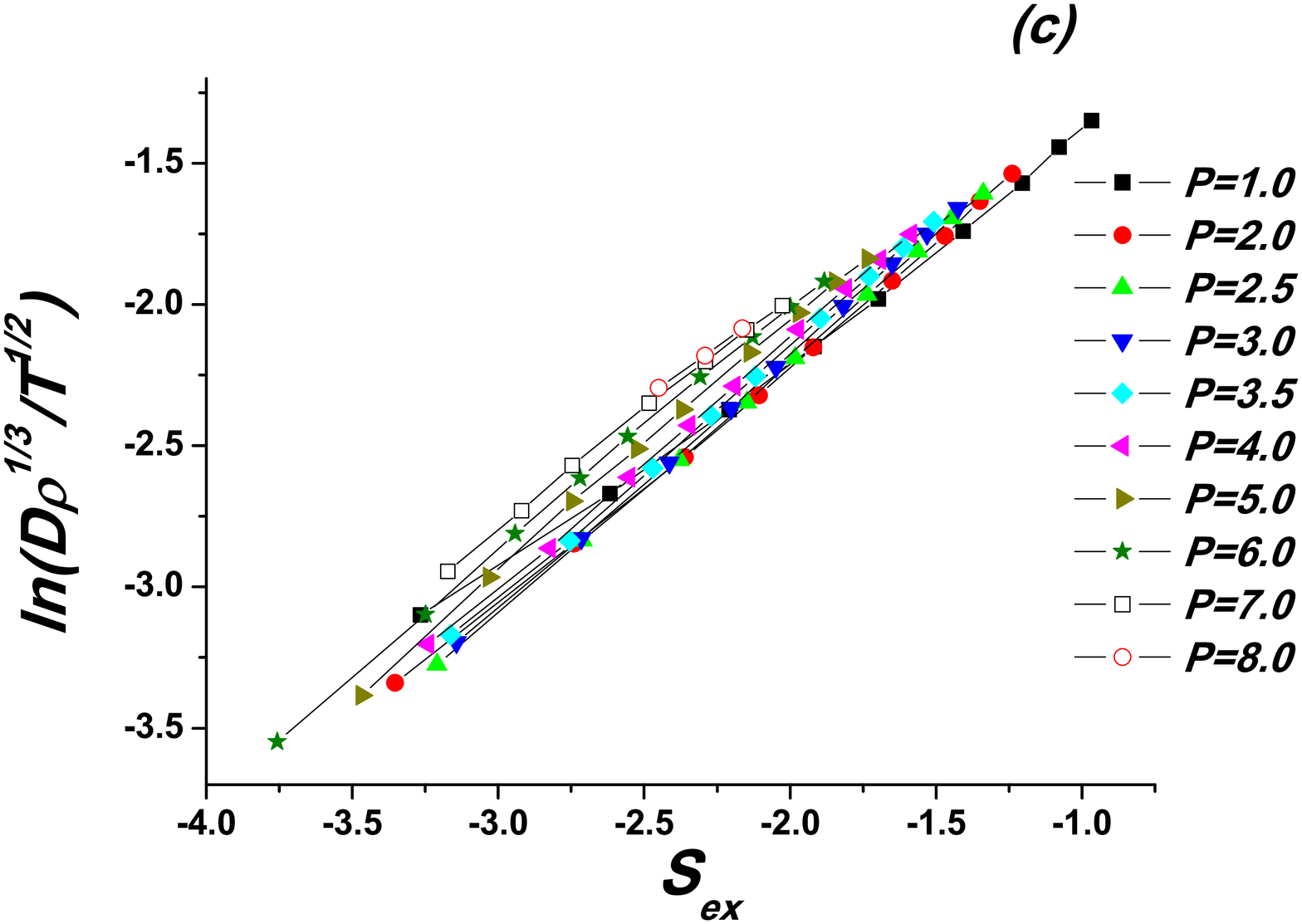}%

\caption{\label{fig:fig9} (Color online). Rosenfeld relation for
SRSS along (a) isochors, (b) isotherms, and (c) isobars.}
\end{figure}

However, from this speculation it follows that the Rosenfeld
scaling should hold true along the trajectories which do not
contain anomalies, i.e. isochors and isobars. In our recent
publication \cite{werostr} we considered the Rosenfeld relation
along isotherms and isochors. Here we bring these trajectories for
the sake of completeness and add the verification of the Rosenfeld
relation along isobars for the purely repulsive potential
(\ref{2}) (Figs. ~\ref{fig:fig9}(a) - (c)). One can see that the
Rosenfeld relation does break down along isotherms which is
consistent with the speculation above. At the same time it holds
true along both isochors and isobars which is consistent with the
monotonous behavior of both diffusion coefficient and excess
entropy along these trajectories.

\begin{figure}
\includegraphics[width=7cm, height=7cm]{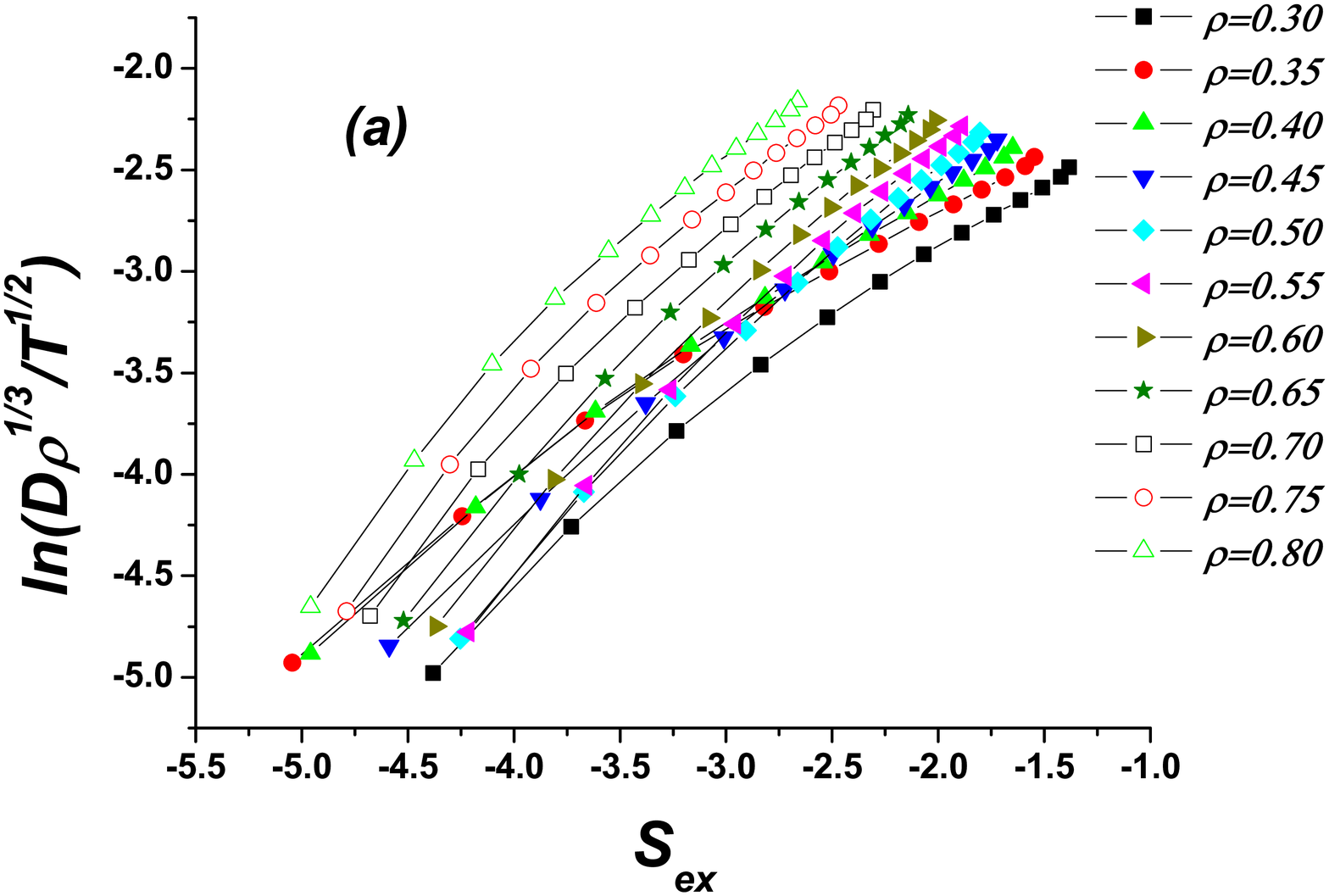}%

\includegraphics[width=7cm, height=7cm]{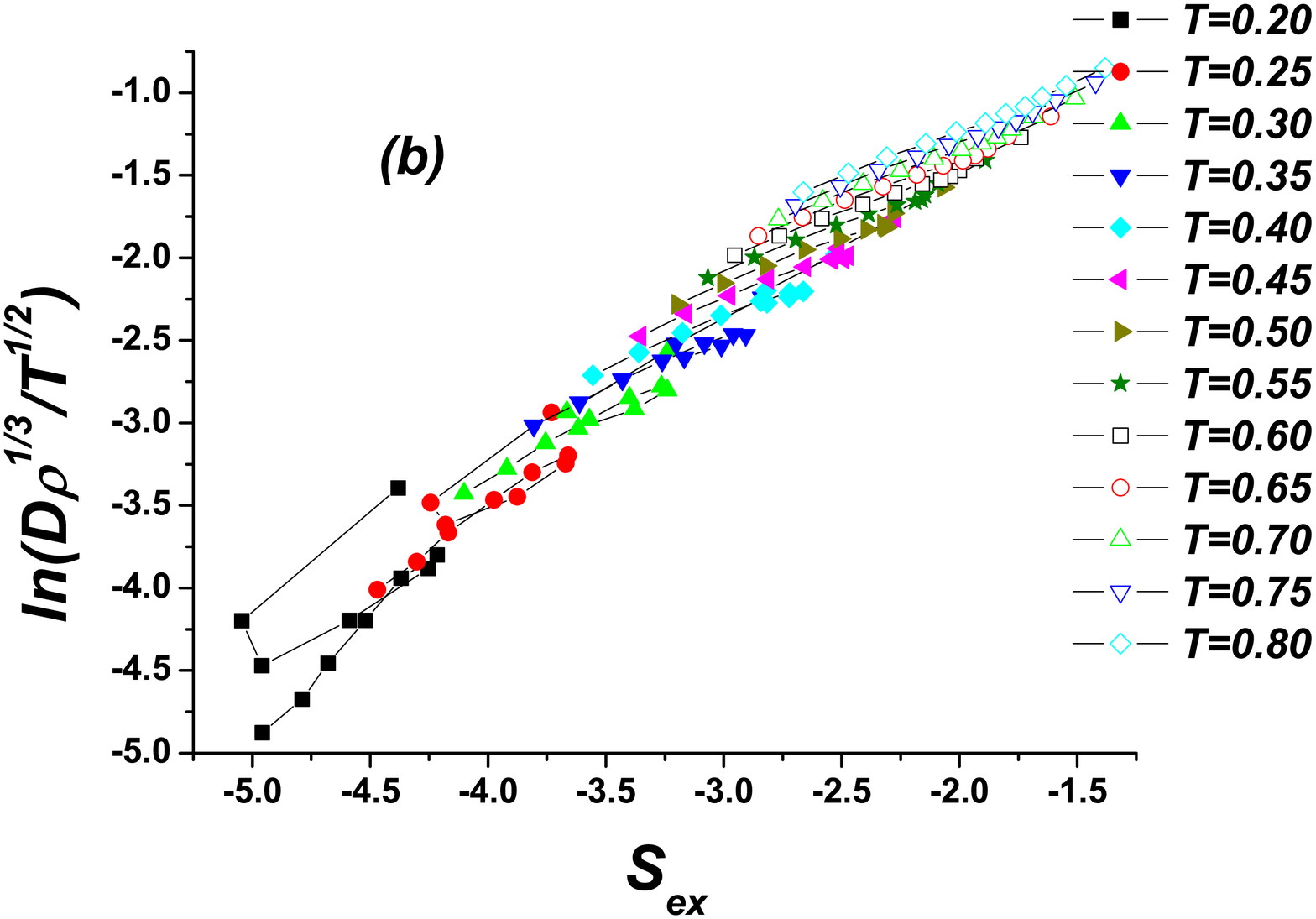}%

\includegraphics[width=7cm, height=7cm]{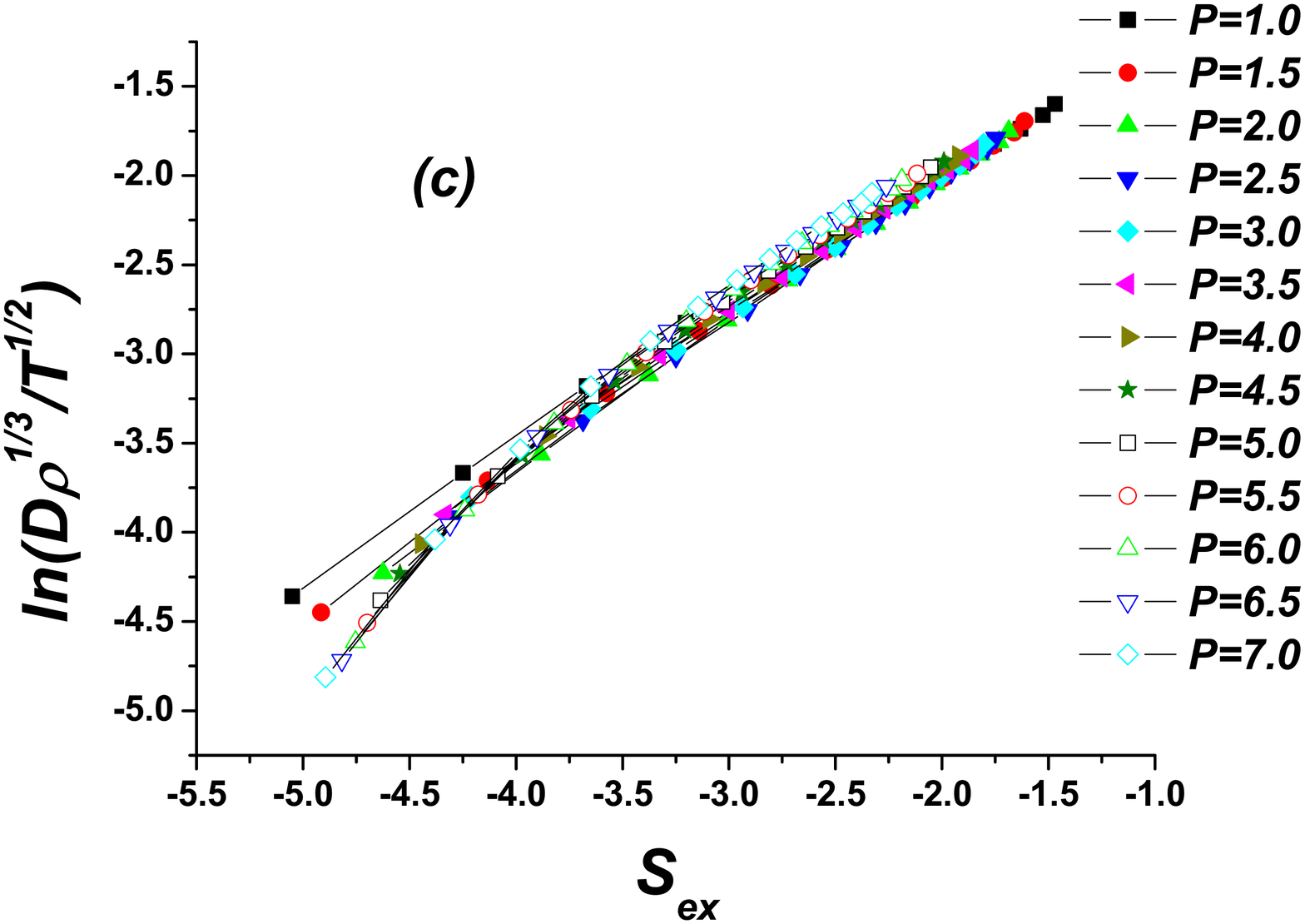}%

\caption{\label{fig:fig19} (Color online). Rosenfeld relation for
SRSS-AW along (a) isochors, (b) isotherms, and (c) isobars.}
\end{figure}

In Figs. ~\ref{fig:fig19}((a) - (c)) we show the Rosenfeld
relation for the system 3. One can see that the behavior is the
same as in the case of the purely repulsive potential.

This observation makes evident that Rosenfeld relation is valid
only along the trajectories without anomalous behavior.

\section{V. Summary and Discussion}

To summarize, in the present article we carry out a molecular
dynamics study of the core-softened systems (SRSS and SRSS-AW) and
show that the anomalous behavior can be seen only along some
particular trajectories in $(P,\rho,T)$ space along which the
behavior of the system is studied.  For example, diffusion and
structural anomalies are visible along isotherms as a function of
density, but disappears along the isochores and isobars as a
function of temperature. On the other hand, the diffusion anomaly
may be seen along adiabats  as a function of temperature, density
and pressure. Density anomaly exists along isochors, isobars and
adiabats. However, if a single curve does not demonstrate the
anomalous behavior, having a set of the curves, one can see the
presence of anomalies via the curves crossing.

We also analyze the applicability of the Rosenfeld entropy scaling
relations to this system in the regions with the water-like
anomalies. It is shown that the validity of the Rosenfeld scaling
relation for the diffusion coefficient also depends on the
trajectory in the $P-\rho-T$ space along which the kinetic
coefficients and the excess entropy are calculated. In particular,
it is valid along isochors and isobars, but it breaks down along
isotherms. The breakdown of the Rosenfeld relation along isotherms
is related to the fact that the boundaries of different anomalies
do not coincide with each other.

The influence  of attraction on the diffusion anomaly is
discussed. It is shown that the attraction makes the anomalies
much more clear and may be found even in the cases when the
anomalies are hardly seen for the purely repulsive potential due
to simulation limitations.

The question is whether this type of behavior is universal for all
types of systems demonstrating the anomalous behavior. It is now
widely believed that the water anomalies are related with the
hypothesized liquid-liquid critical point, the terminal point of a
line of first-order liquid-liquid phase transitions
\cite{stanley1, stanley2,poole,PNAS1,PNAS2}. The anomalies arise
from crossing the Widom line emanating from the hypothesized
liquid-liquid critical point (LLCP) \cite{PNAS1,PNAS2,widom1}. In
particular, it was shown \cite{PNAS1} that the dynamical crossover
from Arrhenius to non-Arrhenius behavior (strong-fragile
transition) takes place both for water and Jagla \cite{17} model.
The Jagla model with attraction displays (without the need to
supercool) a liquid-liquid coexistence line that, unlike water,
has a positive slope \cite{PNAS1}. We believe that our model
should show the similar phase behavior as the Jagla one, however,
with the LLCP in the deeply supercooled region \cite{FRT2011}.

\begin{figure}
\includegraphics[width=7cm, height=7cm]{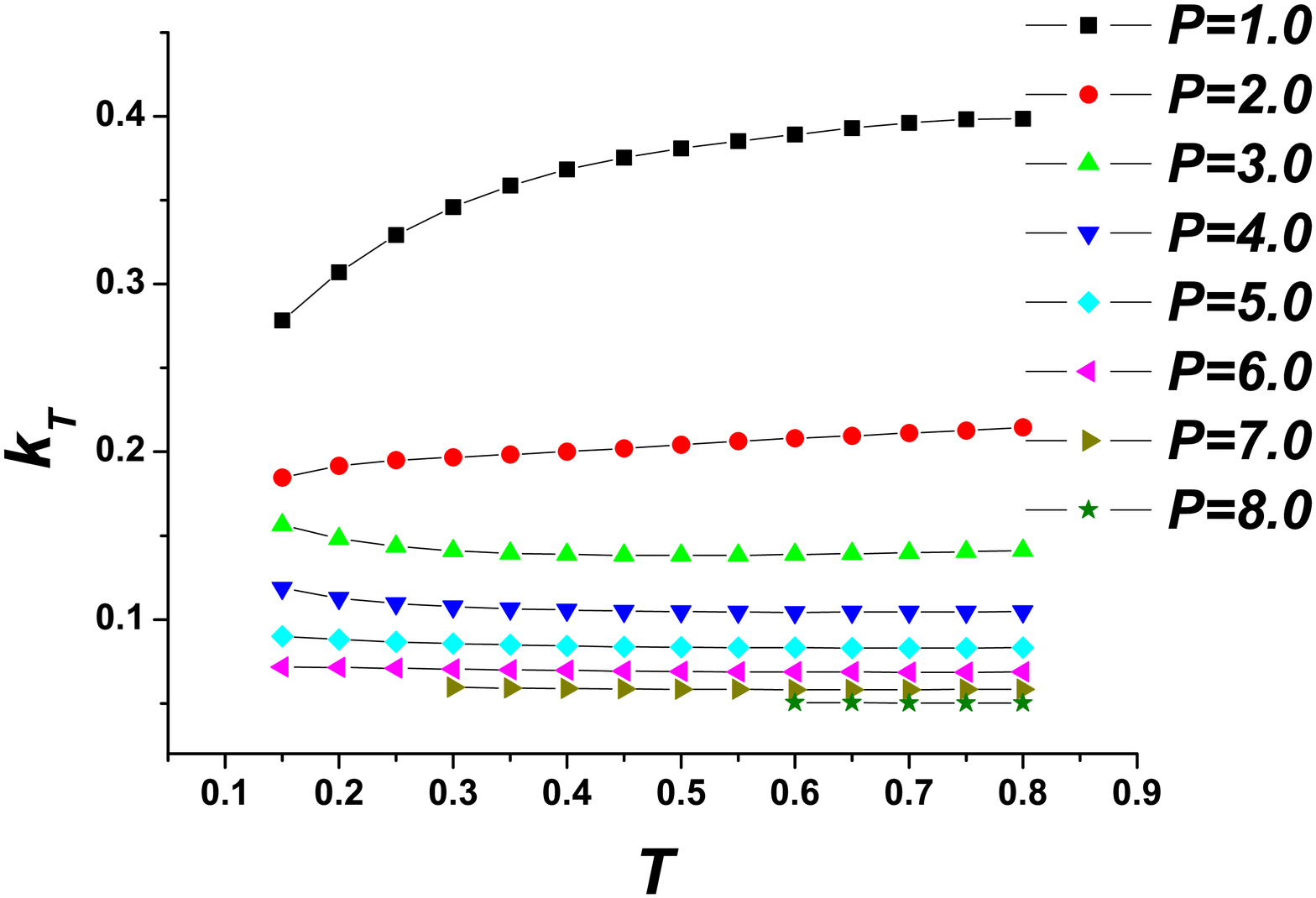}%

\includegraphics[width=7cm, height=7cm]{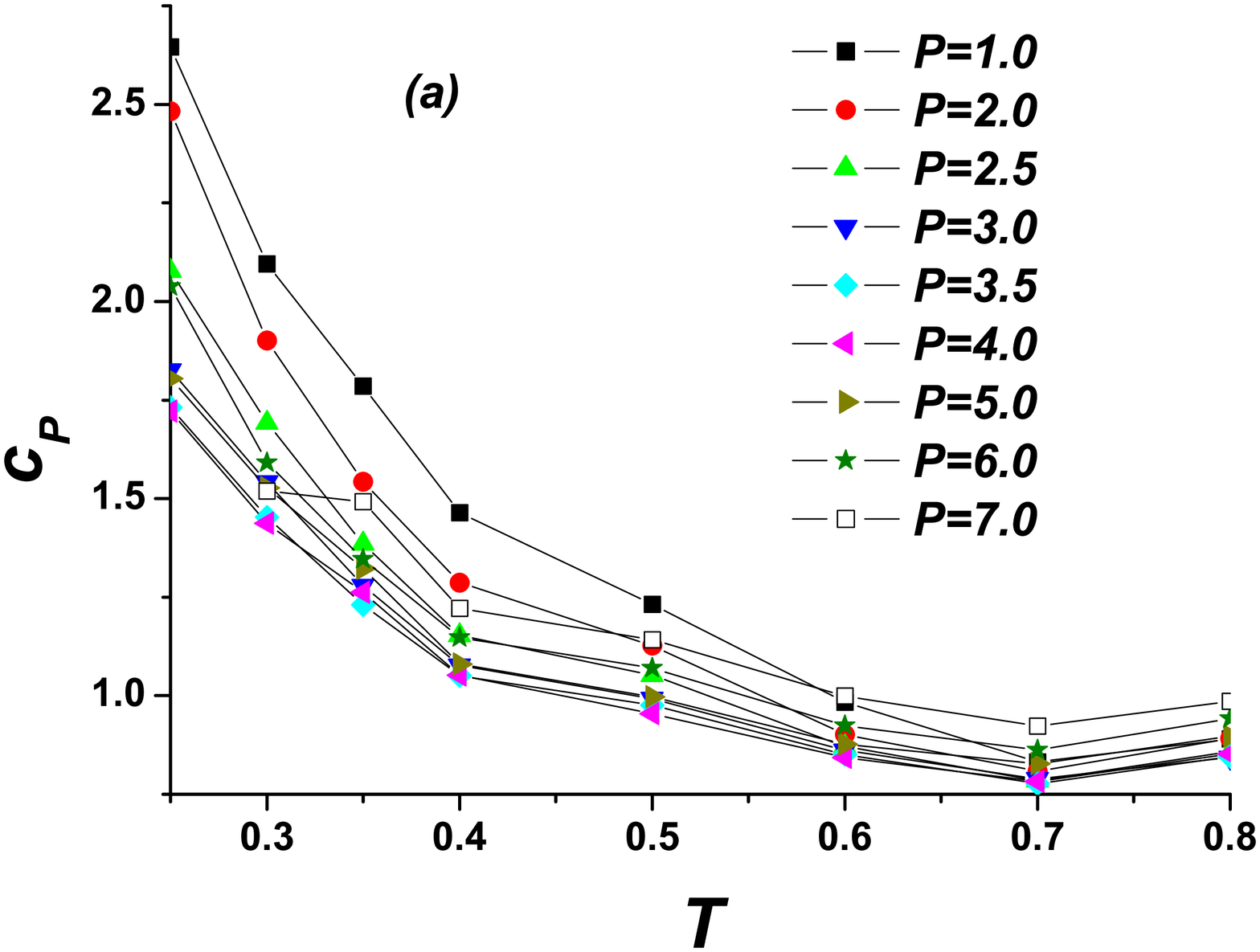}%

\includegraphics[width=7cm, height=7cm]{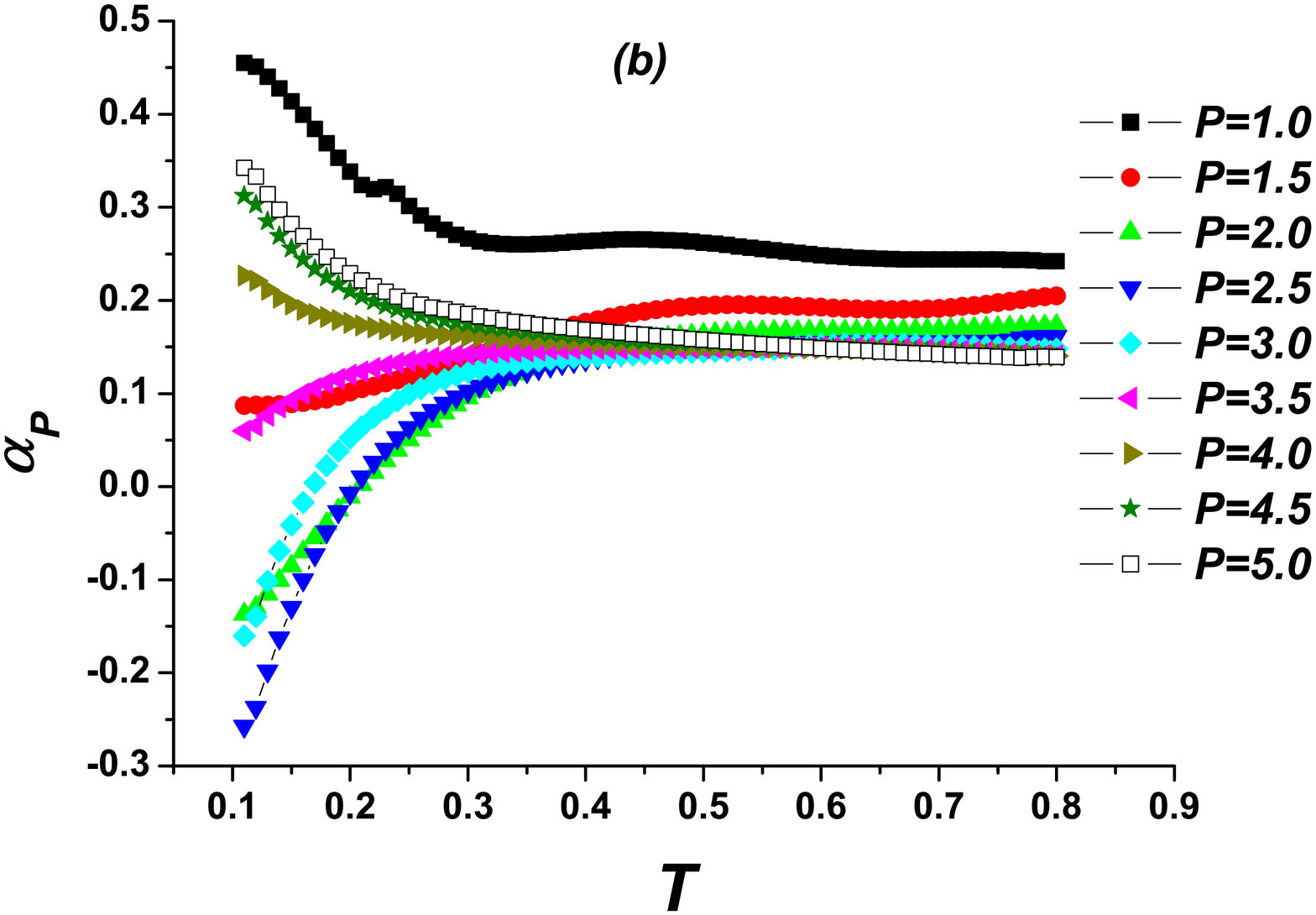}%

\caption{\label{fig:fig20} (Color online). Isothermal
compressibility $K_T$ (a), isobaric heat capacity $C_P$ (b), and
thermal expansion coefficient $\alpha_P$ (c) as a function of
temperature along isobars for the system 1 in Table 1.}
\end{figure}

\begin{figure}
\includegraphics[width=7cm, height=7cm]{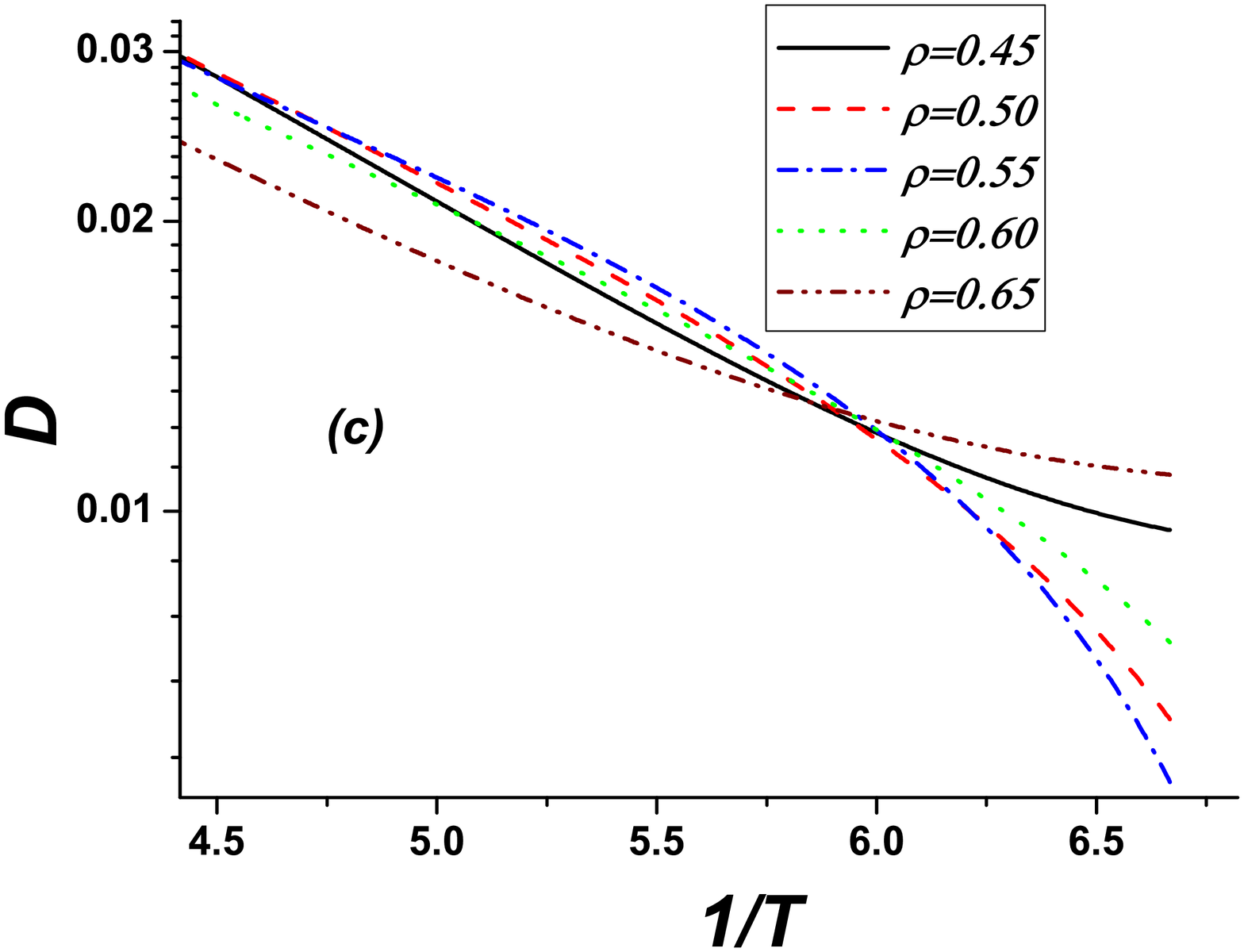}%

\caption{\label{fig:fig21} (Color online). Diffusivity $D$ along
isochores for the system 1 in Table 1.}
\end{figure}

For example, in  Figs.~\ref{fig:fig20} ((a) - (c)) we show the
behavior of the isothermal compressibility $K_T$, the isobaric
heat capacity $C_P$, and the thermal expansion coefficient
$\alpha_P$ for the system 1 in Table 1. One can see that the
system demonstrates the behavior compatible with hypothesis of the
liquid-liquid critical point in the deep supercooled region of the
phase diagram below the homogeneous nucleation line (compare, for
example, with Fig. 1 in Ref. \cite{widom1}) . In
Fig.~\ref{fig:fig21} one can see the dynamical crossover from
Arrhenius to non-Arrhenius behavior. At high temperature, $D$
exhibits an Arrhenius behavior, whereas at low temperature it
follows a non-Arrhenius one. From Figs.~\ref{fig:fig20} and
\ref{fig:fig21} one can easily see that the anomalous lines
correspond to the regions of anomalies in Fig.~\ref{fig:fig1a}
(b). This result is consistent with the case of the Jagla model
\cite{PNAS1}( see Fig. 5 (c) in Ref. \cite{PNAS1}), so we believe
that in our case the Widom line should have the same slope, as in
this model, and we can expect that the anomalous behavior
discussed in this article will be the same for the Jagla model
too, however, the case of water needs an additional investigation.

\bigskip

\begin{acknowledgments}
We thank V. V. Brazhkin for stimulating discussions. Y.F. and E.T.
also thank Russian Scientific Center Kurchatov Institute and Joint
Supercomputing Center of Russian Academy of Science for
computational facilities. The work was supported in part by the
Russian Foundation for Basic Research (Grants No 10-02-00694a,
10-02-00700 and 11-02-00-341a) and Russian Federal Programs
02.740.11.5160 and 02.740.11.0432.
\end{acknowledgments}

\end{document}